# Syntactic Markovian Bisimulation for Chemical Reaction Networks

Luca Cardelli[1], Mirco Tribastone[2], Max Tschaikowski[2], and Andrea Vandin[2]

[1] Microsoft Research & University of Oxford, UK
[2] IMT School for Advanced Studies Lucca, Italy

**Abstract.** In chemical reaction networks (CRNs) with stochastic semantics based on continuous-time Markov chains (CTMCs), the typically large populations of species cause combinatorially large state spaces. This makes the analysis very difficult in practice and represents the major bottleneck for the applicability of minimization techniques based, for instance, on lumpability. In this paper we present syntactic Markovian bisimulation (SMB), a notion of bisimulation developed in the Larsen-Skou style of probabilistic bisimulation, defined over the structure of a CRN rather than over its underlying CTMC. SMB identifies a lumpable partition of the CTMC state space a priori, in the sense that it is an equivalence relation over species implying that two CTMC states are lumpable when they are invariant with respect to the total population of species within the same equivalence class. We develop an efficient partition-refinement algorithm which computes the largest SMB of a CRN in polynomial time in the number of species and reactions. We also provide an algorithm for obtaining a quotient network from an SMB that induces the lumped CTMC directly, thus avoiding the generation of the state space of the original CRN altogether. In practice, we show that SMB allows significant reductions in a number of models from the literature. Finally, we study SMB with respect to the deterministic semantics of CRNs based on ordinary differential equations (ODEs), where each equation gives the time-course evolution of the concentration of a species. SMB implies forward CRN bisimulation, a recently developed behavioral notion of equivalence for the ODE semantics, in an analogous sense: it yields a smaller ODE system that keeps track of the sums of the solutions for equivalent species.

## 1 Introduction

Chemical reaction networks (CRNs) are a powerful model of interaction at the basis of many branches of science such as organic and inorganic chemistry, ecology, epidemiology and systems biology. In computer science, the interpretation of biological systems as computing devices has stimulated a vigorous line of research ranging from the understanding of the computational power of such models (e.g., [48, 41, 13]) to the development of formal techniques for their specification, analysis, and verification (e.g., [15, 30, 17, 46]).

Traditionally, CRNs have been equipped with the well-known quantitative semantics based on a system of ordinary differential equations (ODEs), where an ODE relates to the time-course deterministic evolution of the concentration of each species. It is well-known, however, that such semantics may not always accurately reflect the observed behavior, for example when some species are present in low copies [22]. The alternative stochastic semantics based on continuous-time Markov chains (CTMCs) may provide more accurate estimates, but at an increased computational expense. Indeed, since each CTMC state is a population vector giving the number of copies of each species, there is a combinatorial explosion of the CTMC state space as a function of the initial population of species. In order to cope with this, it would be highly desirable to be able to perform CTMC aggregation, e.g., based on lumpability [5, 20]. However, its applicability in practice is fundamentally hampered by the fact that available methods require to explicitly enumerate the state space. This is typically infeasible for realistic CRNs sizes, or even impossible because CRNs may give rise to infinite state spaces.

Inspired from the seminal work of Larsen and Skou on probabilistic bisimulation [37], in this paper we propose a reduction technique that avoids the generation of the original state space. Instead of reasoning at the semantic level, we identify conditions on the *CRN syntax*. More precisely, we provide a new notion of equivalence over CRN species, called *syntactic Markovian bisimulation* (SMB), based on properties that can be checked by inspecting the set of reactions only, but it induces a partition on states of the underlying CTMC: two CTMC states are related if they are invariant with respect to the total population of species in the same SMB equivalence class. To clarify this, suppose we have a CRN with species $A$, $B$, and $C$, and the SMB that gives the partition $\{\{A, B\}, \{C\}\}$. Then the CTMC state $(n_A = 1, n_B = 2, n_C = 1)$ belongs to the same block as state $(n_A = 2, n_B = 1, n_C = 1)$ because they have equal sums within the equivalence classes. The resulting CTMC partition is an ordinarily lumpable one [5]: in the lumped CTMC each macro-state represents the sum of the probabilities of the original states of a partition block.

Importantly, the lumped CTMC can be obtained avoiding the generation of the original state space altogether, owing to an algorithm that constructs a quotient CRN for an SMB. The possibility of such a CRN-to-CRN transformation is useful not only for model *minimization*, but also for using bisimulation as a technique for model *comparison*. This has received increased attention, largely motivated by applications to evolutionary biology [26, 7, 8, 14, 11, 9, 47, 43].

SMB turns out to be a natural extension of the ordinary lumpability condition (defined on the underlying CTMC semantics) to the CRN syntax. Ordinary lumpability relates two CTMC states whenever they have the same cumulative transition rates toward any partition block. Analogously, SMB relates two species when, roughly speaking, the cumulative kinetic parameters of the reactions where they are involved as reagents are the same for every *lifted equivalence class* of products. This lifting is defined by relating two products that are invariant up to the SMB equivalence classes, as above. An important consequence of this definition style is that it allows us to also extend the aforementioned CTMC minimization algorithms to SMB. In particular, we present an algorithm for computing the largest SMB that refines a given input partition of species in polynomial time and space.

Being syntactically driven, it is perhaps not surprising that SMB is only a sufficient condition for CTMC lumpability. As a consequence, it is important to understand to what extent it can be effectively applied in practice. On CRN models of biological systems taken from the literature we show that SMB can achieve substantial compressions, yielding reduced CRNs with significantly fewer species and reactions in some cases. We measure the impact of SMB on the analysis of the CRN when this is done by means of stochastic simulation [28], the method of choice in realistic systems due to the large state spaces involved (e.g., [18]). We report noticeable runtime speed-ups in many cases, up to two orders of magnitude, even allowing the execution of benchmark models that would otherwise generate out of memory errors if not reduced. These numerical tests also reveal an interesting connection between SMB and the deterministic semantics of CRNs: the equivalence classes of species found in all the analyzed models coincide with those recently reported in [8] for *forward CRN bisimulation* (FB), an equivalence relation over species that aggregates related ODEs in an analogous way, exactly preserving their total concentration trajectories at all time points. We explain this fact by showing that SMB implies an FB, however the converse is not true in general. Nevertheless, in our tests FB was not able to aggregate more than SMB.

**Further related work.** The closest approach to ours is by Feret et al. [25] who identify *stochastic fragments* on the rule-based language $\kappa$ [19]. These represent syntactic criteria that yield a sufficient condition for *weak* lumpability (see, e.g., [5]) on the CTMC. The advantage is that the rule-based model is often combinatorially smaller than its underlying CRN description; however, the approach is domain specific in that it can be applied to systems, e.g., protein-protein interaction networks, which can be conveniently expressed as rule-based systems. On the contrary, since SMB works at the level of the CRN it is more general, at the expense of a more expensive syntactic analysis in this application domain.

For process algebra with quantitative semantics based on CTMCs, several approaches have been proposed for *on-the-fly* computations of lumped chain that avoid the generation of the original state space. These are based on deriving transitions of the lumped chain from a canonical representative of an equivalence class (e.g., [31, 29, 36, 45]). Here considerable state-space compressions are owed to *symmetry reduction*, whereby identical copies of a process in parallel composition can be collapsed through a lumpable partition that contains all processes that are equal up to a permutation of the composed sub-terms. Symmetry reduction could be useful in the case that the CRN is described at the individual molecular level, as for instance in Cardelli's Chemical Ground Form [6]. However, we remark that a CRN gives a CTMC that tracks the population sizes of each species, implicitly accounting already for symmetry due to the assumption that two molecules of the same species are identical. SMB, instead, captures structural relations, see [8] for a physical interpretation of some equivalence classes. In this sense, SMB is closer in spirit to the idea of *place bisimulation* for Petri nets, which establishes a relation over places that induces a bisimulation in the classical, non-quantitative strong sense [2].

**Paper structure.** Section 2 introduces the notion of CRN and defines its semantics. SMB is introduced in Section 3, while in Section 4 it is shown that SMB induces a reduced CRN whose CTMC is related via ordinary lumpability to the CTMC of the original CRN. Section 5 presents the algorithm for computing the largest SMB. Applicability and efficiency of the algorithm are demonstrated on biological models from the literature in Section 6. A formal comparison of SMB with FB complements the experiments. Conclusions are drawn in Section 7.

## 2 Chemical Reaction Networks

In this paper we consider *mass-action* CRNs, where each reaction is labeled with a constant, the reaction rate. The speed of the reaction will be proportional with this rate to the product of the abundances of the reactants. In particular, we focus on basic chemistry where only *elementary reactions* are considered: *unary reactions*, involving a single reactant performing a spontaneous reaction, and *binary reactions*, where two reactants interact; we call a binary reaction a *homeoreaction* if the two reactants are of the same species. Elementary reactions pose no restrictions on products. Several models found in the literature (including those discussed in Section 6) belong to this class. Also, this is consistent with the physical considerations which stipulate that reactions with more than two reactants are very unlikely to occur in nature [27]. In the rest of the paper we will refer to such *elementary mass-action CRNs* as just CRNs.

Formally, a CRN $(S, R)$ is a set of species $S$ and a set of chemical reactions $R$. Each reaction is a triple written in the form $\rho \xrightarrow{\alpha} \pi$, where $\rho$ and $\pi$ are the multi-sets of species representing the *reactants* and *products*, respectively, and $\alpha \geq 0$ is the reaction rate. We denote by $\rho(X)$ the multiplicity of species $X$ in the multi-set $\rho$, and by $\mathcal{MS}(S)$ the set of finite multi-sets of species in $S$. To adhere to standard chemical notation, we shall also use the operator $+$ to denote multi-set union, e.g., $X + Y + Y$ (or just $X + 2Y$) denotes the multi-set of species $\{\!|X, Y, Y|\!\}$; similarly $\rho - X$ denotes multi-set difference $\rho \setminus \{\!|X|\!\}$. We also use $X$ to denote either the species $X$ or the singleton $\{\!|X|\!\}$.

*Example 1.* We now provide a simple CRN, $(S_e, R_e)$, with $S_e = \{A, B, C, D, E\}$ and

$$R_e = \{A \xrightarrow{6} D, A \xrightarrow{2} 3C, C+D \xrightarrow{5} 2C+D, B \xrightarrow{6} C,$$
$$B \xrightarrow{2} 3D, E+D \xrightarrow{5} 2C+D, 2D \xrightarrow{3} C\},$$

which will be used as a running example throughout the paper.

We next recall the well-known CTMC semantics of CRNs (see, e.g., [6, 28]), which allows us to associate a population-based CTMC to a given CRN and an initial population of its species. Here the state descriptor gives the number of elements for each species, hence it is formally represented as a multi-set of species. The CTMC specification is mediated by a multi-transition system (MTS), to record multiplicity of transitions. This is needed to account for two or more reactions contributing to the same CTMC transition, e.g., $A + B \xrightarrow{\alpha_1} B + C$ and $A \xrightarrow{\alpha_2} C$. The whole state space is defined by enumerating states, starting from some initial state.

**Definition 1 (Multi-transition system of a CRN).** *Let $(S, R)$ be a CRN. The multiset of outgoing transitions from state $\sigma \in \mathcal{MS}(S)$ is obtained as*

$$out(\sigma) = \{\!|\sigma \xrightarrow{\alpha \cdot \sigma(X)} \sigma - X + \pi \mid (X \xrightarrow{\alpha} \pi) \in R |\!\}$$
$$\uplus \{\!|\sigma \xrightarrow{\alpha \cdot \sigma(X) \cdot \sigma(Y)} ((\sigma - X) - Y) + \pi \mid X \neq Y \wedge (X + Y \xrightarrow{\alpha} \pi) \in R |\!\}$$
$$\uplus \{\!|\sigma \xrightarrow{\frac{\alpha}{2} \cdot \sigma(X) \cdot (\sigma(X)-1)} ((\sigma - X) - X) + \pi \mid (X + X \xrightarrow{\alpha} \pi) \in R |\!\}$$

*The set of reachable states from $\sigma$, denoted by $reach(\sigma)$, is the smallest set such that: i) $\sigma \in reach(\sigma)$; and (ii) if $\sigma' \in reach(\sigma)$, then the target states of $out(\sigma')$ belong to $reach(\sigma)$. Finally, for an initial state $\sigma_0 \in \mathcal{MS}(S)$, the MTS for $(S, R)$ and $\sigma_0$ is the union of the multi-sets of transitions outgoing from any reachable state, i.e. $MTS(\sigma_0) = \biguplus_{\theta \in reach(\sigma_0)} out(\theta)$.*

We note that each reaction $\rho \xrightarrow{\alpha} \pi$ can be applied to source states $\sigma$ containing $\rho$, i.e. $\sigma = \sigma' + \rho$ for some multi-set $\sigma'$. The corresponding target state is $\sigma' + \pi$. The rate for unary reactions $X \xrightarrow{\alpha} \pi$ is $\alpha \cdot \sigma(X)$ and accounts for the fact that each instance of the reagent can perform the reaction independently. For binary reactions $X + Y \xrightarrow{\alpha} \pi$ with $X \neq Y$, instead, the transition rate is proportional to the product of the populations of the species involved, i.e. $\alpha \cdot \sigma(X) \cdot \sigma(Y)$. This corresponds to the number of possible interactions between molecules, proportionally to the reaction propensity $\alpha$ [35, 28]. For a homeoreaction involving $X$, the number of distinct interactions is given by $\binom{\sigma(X)}{2} = \frac{1}{2} \cdot \sigma(X) \cdot (\sigma(X) - 1)$.

*Example 2.* Consider the initial population $\sigma_{0e} = 2A + C + D$ for $(S_e, R_e)$. Then we have $out(\sigma_{0e}) = \{\!|\sigma_{0e} \xrightarrow{6 \cdot 2} A + C + 2D, \sigma_{0e} \xrightarrow{2 \cdot 2} A + 4C + D, \sigma_{0e} \xrightarrow{5} 2A + 2C + D|\!\}$. The three transitions are due, respectively, to the first, second and third reaction of $R_e$.

We wish to stress the difference between a CRN and its MTS. While both are collections of triples in $\mathcal{MS}(S) \times \mathbb{R} \times \mathcal{MS}(S)$, the elements of the former are *syntactic*. Instead, the nature of the latter is *semantic* because it induces the underlying CTMC. In particular, given an MTS the CTMC is obtained by collapsing all transitions between the same source and target into a single CTMC transition and summing their rates.

**Definition 2 (CTMC semantics).** *Let $(S, R)$ be a CRN, and $\sigma_0$ an initial population. The CTMC of $(S, R)$ for $\sigma_0$ has states $reach(\sigma_0)$ and its transitions are given by*

$$MC(\sigma_0) = \{\sigma \xrightarrow{r} \theta \mid \sigma, \theta \in reach(\sigma_0) \wedge \sigma \neq \theta \wedge r = \sum_{\sigma \xrightarrow{r'} \theta \in MTS(\sigma_0)} r' \}.$$

*For any two states $\sigma, \theta \in \mathcal{MS}(S)$ the element of the* infinitesimal generator matrix *of $MC(\sigma_0)$ from $\sigma$ to $\theta$ is defined as:*

$$q(\sigma, \theta) = \begin{cases} r & \text{if } \sigma \neq \theta \wedge \sigma \xrightarrow{r} \theta \in MC(\nu_0) \\ -\sum_{\theta' \in \mathcal{MS}(S) \text{ s.t. } \theta' \neq \sigma} q(\sigma, \theta') & \text{if } \sigma = \theta \\ 0 & otherwise \end{cases}$$

*For any $\mathcal{M} \subseteq \mathcal{MS}(S)$, we define $q[\sigma, \mathcal{M}] = \sum_{\theta \in \mathcal{M}} q(\sigma, \theta)$ and $q[\mathcal{M}, \theta] = \sum_{\sigma \in \mathcal{M}} q(\sigma, \theta)$.*

## 3 Syntactic Markovian Bisimulation

This section introduces Syntactic Markovian Bisimulation (SMB) as a sufficient condition for CTMC ordinary lumpability. We first recast this latter notion to our notation.

**Definition 3.** *Let $(S, R)$ be a CRN, $\sigma_0$ an initial population, $MC(\sigma_0)$ the underlying CTMC and $\mathcal{H}$ a partition of $\mathcal{MS}(S)$. Then, $MC(\sigma_0)$ is* ordinarily lumpable *with respect to $\mathcal{H}$ iff for any $\sigma_1, \sigma_2$ in the same block of $\mathcal{H}$ we have $q[\sigma_1, \mathcal{M}] = q[\sigma_2, \mathcal{M}]$ for all $\mathcal{M} \in \mathcal{H}$.*

Lumpability is given in terms of an equivalence relation among the states of a CTMC. Instead, SMB is an equivalence over the species of a CRN. Note that there is no one-to-one correspondence between species and the state space of the CTMC underlying a CRN. Indeed, the species define the state descriptor, but the cardinality of the state space is typically much larger since it depends on all the possible configurations of populations that are reachable from a given initial population. Thus we need to lift a relation over species to one over CTMC states. We do so by providing the notion of *multi-set lifting*: given a CRN $(S, R)$ and an equivalence relation $\mathcal{R}$ over $S$, the lifting of $\mathcal{R}$ relates multi-sets with same number of $\mathcal{R}$-equivalent species.

**Definition 4 (Multi-set Lifting).** *Let $(S, R)$ be a CRN, $\mathcal{R} \subseteq S \times S$ be an equivalence relation over $S$, and $\mathcal{H}$ be the partition induced by $\mathcal{R}$ over $S$. We define the* multi-set lifting *of $\mathcal{R}$ on $\mathcal{MS}(S)$, denoted by $\mathcal{R}^\uparrow \subseteq \mathcal{MS}(S) \times \mathcal{MS}(S)$, as*

$$\mathcal{R}^\uparrow \triangleq \{(\sigma_1, \sigma_2) \mid \sigma_1, \sigma_2 \in \mathcal{MS}(S) \land \forall H \in \mathcal{H} : \sum_{X \in H} \sigma_1(X) = \sum_{X \in H} \sigma_2(X)\}$$

The multi-set lifting of $\mathcal{R}$ can be readily seen to be an equivalence relation over $\mathcal{MS}(S)$.

*Example 3.* Consider the equivalence relation $\mathcal{R}_m$ over $S_e$ inducing $\mathcal{H}_m = \{\{A\}, \{B\}, \{C, E\}, \{D\}\}$. Examples of multi-sets related by $\mathcal{R}_m^\uparrow$ are $C$ and $E$, $2C$ and $2E$, and $C + E$ and $2E$, while $(A + C, B + C) \notin \mathcal{R}_m^\uparrow$.

The syntactic checks of SMB are performed via the notion of *reaction rate* given below. It computes, in essence, the cumulative rate that transforms a given reagent $\rho$ into a certain product $\pi$.

**Definition 5 (Reaction rate).** *Let $(S, R)$ be a CRN, and $\rho, \pi \in \mathcal{MS}(S)$. The reaction rate from $\rho$ to $\pi$ is defined as $\mathbf{rr}(\rho, \pi) = \sum_{\rho \xrightarrow{\alpha} \pi \in R} \alpha$. For any $\mathcal{M} \subseteq \mathcal{MS}(S)$, we define $\mathbf{rr}[\rho, \mathcal{M}] = \sum_{\pi \in \mathcal{M}} \mathbf{rr}(\rho, \pi)$.*

We can now define SMB.

**Definition 6 (Syntactic Markovian Bisimulation).** *Let $(S, R)$ be a CRN, $\mathcal{R}$ an equivalence relation over $S$, $\mathcal{R}^\uparrow$ the multi-set lifting of $\mathcal{R}$ and $\mathcal{H}^\uparrow = \mathcal{MS}(S)/\mathcal{R}^\uparrow$. We say that $\mathcal{R}$ is a syntactic Markovian bisimulation (SMB) for $(S, R)$ if and only if*

$$\mathbf{rr}[X + \rho, \mathcal{M}] = \mathbf{rr}[Y + \rho, \mathcal{M}], \textit{for all } (X, Y) \in \mathcal{R}, \rho \in \mathcal{MS}(S), \textit{and } \mathcal{M} \in \mathcal{H}^\uparrow.$$

*We define the syntactic Markovian bisimilarity of $(S, R)$ as the union of all SMBs of $(S, R)$.*

*Remark 1.* Note that the multi-sets $X + \rho$ and $Y + \rho$ differ only in one species ($X$ and $Y$), thus projecting comparisons involving multisets (i.e., $X + \rho$ and $Y + \rho$) onto species (i.e., $X$ and $Y$). In this view, $\rho$ plays a role similar to an action type in traditional bisimulations, since it restricts interactions with a given reagent partner (or $\emptyset$ in case of unary reactions). Furthermore, Definition 6 entails a finite number of checks because all evaluations of **rr** are equal to zero for multisets that are not products in the CRN, see the algorithm of Section 5.

*Example 4.* Consider again $\mathcal{R}_m$ and $\mathcal{H}_m$. From $\mathbf{rr}(C+D, 2C+D) = \mathbf{rr}(E+D, 2C+D)$ can be inferred that $\mathcal{R}_m$ is an SMB.

As usual, we are interested in the largest bisimulation. The next result ensures that syntactic Markovian bisimilarity is an SMB, thus showing that it is also the largest one. Following the approach of [32], we show this by proving that the transitive closure of a union of SMBs is an SMB.[1]

**Proposition 1.** *Let $(S, R)$ be a CRN, $I$ a set of indices, and $\mathcal{R}_i$ an SMB for $(S, R)$, for all $i \in I$. The transitive closure of their union $\mathcal{R} = (\bigcup_{i \in I} \mathcal{R}_i)^*$ is an SMB for $(S, R)$.*

We now provide our first major result.

**Theorem 1.** *Let $\mathcal{R}$ be an SMB for the CRN $(S, R)$. Then, its multi-set lifting $\mathcal{R}^\uparrow$ induces the ordinarily lumpable partition $\mathcal{H}^\uparrow$ on $MC(\sigma_0)$ for any initial state $\sigma_0$.*

Three remarks are in order. First, we stress that a single SMB induces infinitely many ordinarily lumpable partitions because there are no constraints on $\sigma_0$. Second, Theorem 1 makes no assumption on the cardinality of the CTMC state space underlying the CRN. In particular, it can also be applied to infinite state spaces; indeed Example 1 is an instance of such a situation because, e.g., of the reaction $C+D \xrightarrow{5} 2C+D$, which may generate infinitely many copies of $C$ whenever the initial state has at least one copy of species $D$ and one of species $C$. The original result of ordinary lumpability applies to finite CTMCs. However, using concepts from functional analysis and the theory of linear ODEs on Banach spaces, this statement can be extended, under certain assumptions, to CTMCs with countably infinite state spaces [39]. A sufficient condition for the theory to apply is to assume that the state space of the CTMC is partitioned in blocks of finite size. Indeed, the multi-set lifting ensures that any CTMC partition stemming from SMB enjoys this property. Third, as anticipated in Section 1, SMB is only a sufficient condition for CTMC ordinary lumpability.

*Example 5.* Consider the CRN $(\{F, G\}, \{F \xrightarrow{\alpha_1} G, G \xrightarrow{\alpha_2} F\})$ with $\alpha_1 \neq \alpha_2$ and $\sigma_0 = F$. The underlying CTMC has the state space $\{F, G\}$ and it readily follows that $\{\{F, G\}\}$ is an ordinarily lumpable partition, while it is not an SMB.

At the same time, however, SMB can be computed efficiently and induces significant reductions to biological models from literature, as discussed in Section 6.

---

[1] The proofs of all statements are in the appendix.

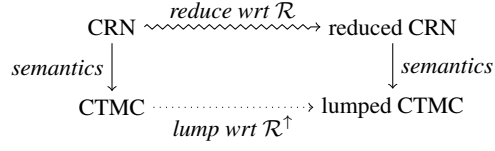

**Fig. 1.** The relation among ($\mathcal{R}$-reduced) CRNs and ($\mathcal{R}^\uparrow$-lumped) semantics, with $\mathcal{R}$ an SMB.

## 4 Reduced CRN

Given a CRN $(S, R)$ and an SMB $\mathcal{R}$, we next provide an algorithm that efficiently computes a $\mathcal{R}$-*reduced CRN* that induces directly the CTMC aggregated according to $\mathcal{R}^\uparrow$, without exploring the state space of the original CTMC. This is visualized in Figure 1.

We wish to point out that this reduction algorithm happens to coincide with the *forward reduction* of [8], which has been applied to obtain a quotient CRN up to an FB, mentioned in Section 1, defined for the ODE semantics of CRNs. For the sake of completeness we state the notion of reduced CRN according to this paper's notation. To this end, we introduce the following notions. Given a partition $\mathcal{H}$ of $S$ such that $\mathcal{H} = S/\mathcal{R}$, let $X^H$ denote the canonical representative of a block $H \in \mathcal{H}$. Moreover, for any $\rho \in \mathcal{MS}(S)$, set $\rho^\mathcal{R} = \sum_{X \in \rho} X^H$ for the multiset obtained replacing each species with its canonical representative. Also, for any $\mathcal{M} \in \mathcal{H}^\uparrow$ we use $\mathcal{M}^\mathcal{R}$ for $\rho^\mathcal{R}$, with $\rho$ any multi-set in $\mathcal{M}$.

**Definition 7 (Reduced CRN).** *Let $(S, R)$ be a CRN, $\mathcal{R}$ an equivalence relation on $S$ and $\mathcal{H} = S/\mathcal{R}$. The $\mathcal{R}$-reduction of $(S, R)$ is defined as $(S, R)^\mathcal{R} = (S^\mathcal{R}, R^\mathcal{R})$, where $S^\mathcal{R} = \{X^H \mid H \in \mathcal{H}\}$ and $R^\mathcal{R}$ is computed as follows: (F1) Discard all reactions $\rho \xrightarrow{\alpha} \pi$ such that $\rho \neq \rho^\mathcal{R}$, i.e. whose reagents have species that are not representatives; (F2) Replace the species in the products of the remaining reactions with their canonical representatives; (F3) Fuse all reactions that have the same reactants and products by summing their rates.*

In the case of our running example, the above definition yields the following.

*Example 6.* Consider the SMB $\mathcal{R}_m$ of Example 3 and the underlying partition $\mathcal{H}_m = \{\{A\}, \{B\}, \{C, E\}, \{D\}\}$. With $C$ being the representative of its block, the $\mathcal{R}_m$-reduction of $(S_e, R_e)$ is $S_e{}^{\mathcal{R}_m} = \{A, B, C, D\}$, $R_e{}^{\mathcal{R}_m} = \{A \xrightarrow{6} D, A \xrightarrow{2} 3C, B \xrightarrow{6} C, B \xrightarrow{2} 3D, C+D \xrightarrow{5} 2C+D, 2D \xrightarrow{3} C\}$. Note that the reaction $E + D \xrightarrow{5} 2C + D$ is discarded.

**Theorem 2.** *Let $(S, R)$ be a CRN, $\mathcal{R}$ denote an SMB and $\mathcal{H} = S/\mathcal{R}$. Further, let $\mathcal{H}^\uparrow$ denote the partition induced by $\mathcal{R}^\uparrow$ on $\mathcal{MS}(S)$. Then, for any initial population $\sigma_0$ of $(S, R)$, the underlying CTMC is such that for all $\sigma \in \mathcal{MS}(S)$ it holds that $q_{(S,R)}[\sigma, \mathcal{M}] = q_{(S^R, R^R)}(\sigma^\mathcal{R}, \mathcal{M}^\mathcal{R})$ for any $\mathcal{M} \in \mathcal{H}^\uparrow$.*

```
1  SMBisimilarity(S, R, H) :=              1  Split(S, R, ρ, M_spl, H, spls) :=
2    H↑ = lift(H, Π(R))                    2    forall (X ∈ S)
3    spls = L(R) × H↑                      3      X.rr0
4    while (spls ≠ ∅)                      4    forall (π ∈ M_spl)
5      (ρ, M_spl) = pop(spls)              5      forall (X + ρ --r--> π ∈ π.inc)
6      Split(S, R, ρ, M_spl, H, spls)      6        X.rr = X.rr + r
                                           7    updatePartitionAndSplitters(S, R, H, spls)
```

**Fig. 2.** Syntactic Markovian bisimilarity

If we assume that each block of an SMB partition stores a pointer to its representative, the reduced CRN can be computed in $O(r \cdot s \cdot \log s)$ steps [8], where $s := |S|$ and $r := |R|$.

## 5 Computing Syntactic Markovian Bisimilarities

Syntactic Markovian bisimilarity can be encoded as a partition refinement problem [38], analogously to well-known algorithms for quantitative extensions of labeled transition systems [33, 3, 10]. Hence, we only detail the conceptually novel parts, i.e., the computation of the quantities in Definition 6 and the notion of multi-set lifting.

**Notation and data structures.** Our algorithm for syntactic Markovian bisimilarity, SMBisimilarity, is given in Figure 2, where $(S, R)$ is the input CRN, and $H$ the initial partition to be refined up to SMB. We use $s := |S|$, $r := |R|$, and $L(R)$ for the set of all *labels*, i.e., all species multi-sets $\rho$ to be considered according to Definition 6 in the computation of **rr**. That is, $L(R) = \{\lbrace\!| X |\!\rbrace \mid X \in S \wedge \exists Y \in S, \exists X + Y \xrightarrow{r} \pi \in R\} \cup \{\lbrace\!| \emptyset |\!\rbrace\}$. We use $\{\lbrace\!| \emptyset |\!\rbrace\}$ to account for unary reactions, and $\lbrace\!| X |\!\rbrace \in L(R)$ for each reagent $X$ occurring in at least one binary reaction in the CRN. We set $l := |L(R)|$, which is bounded by $\min(s + 1, 2 \cdot r)$.

In this pseudo-code we assume that species and reactions are stored in data structures via pointers. Species are stored in a list, while a block of $H$ is a list of its species, each species in turn having a pointer to its block, requiring $O(s)$ space. Also $R$ is stored in a list. Each reaction has two fields for the reagents, and a list of pairs in the form (species,multiplicity) for the products, sorted according to a total ordering on species. Thus, $R$ requires $O(s \cdot r)$ space. Finally, $L(R)$ is stored in a list too, for an overall $O(s \cdot r)$ space complexity.

**Overview.** SMBisimilarity is based on Paige and Tarjan's classical solution to the *relational coarsest partition problem* [38] and quantitative extensions thereof (e.g., [20, 3]). A given initial partition is iteratively refined (i.e., its blocks are split), until a partition satisfying the required conditions is found. Refinements are based on the notion of *splitter*, here given by $(\rho, M_{spl})$, with $\rho \in L(R)$ and $M_{spl} \in H↑$: a block of $H$ is *split* in sub-blocks of species with same $\rho$-reaction rate towards a block $M_{spl}$ of *equivalent* multi-sets of species. We stress that, differently from classic bisimulations, in SMB splitters are blocks of products obtained via the multi-set lifting from a species partition $H$, rather than blocks of $H$ itself. Note that the set $MS(S)$ of all possible multi-sets of species in $S$ is infinite. However, we can restrict to the set $\Pi(R) = \{\pi \mid (\rho \xrightarrow{\alpha} \pi) \in R\}$ only, collecting the multi-sets of species appearing as products in the reactions of the considered CRN. This is because any multi-set in $\mathcal{MS}(S) \setminus \Pi(R)$ will not contribute to the reaction rates. We store $\Pi(R)$ as a list, requiring $O(r)$ space, while a partition of $\Pi(R)$ is encoded by representing a block with a list of pointers to its products.

**The `SMBisimilarity` procedure (Lines 1-6).** The procedure starts (Line 2) by creating the partition $\mathcal{H}^\uparrow$ of $\Pi(R)$ according to the multi-set lifting of $\mathcal{H}$. This requires $O(s \cdot r \cdot \log r)$ time; this is because $O(s \cdot r)$ is required to count the number of species of each block of $\mathcal{H}$, for each product, while $O(s \cdot r \cdot \log r)$ is required to partition the products. This is done by iteratively sorting the products according to the number of $H$-species they have, for each $H \in \mathcal{H}$. It requires $O(r \cdot \log r)$ per block of $\mathcal{H}$. Then, a set `spls` of initial candidate splitters is generated for each $\rho \in \mathcal{L}(R)$ and $\mathcal{M} \in \mathcal{H}^\uparrow$. In order to bound the size of `spls` to $O(s)$ we do not explicitly store each pair $(\rho, \mathcal{M}_{spl})$ for all $\rho$. Instead we store only one, initialized with a reference to the first position of $\mathcal{L}(R)$, and then update the pointer to the next position when necessary.

Lines 4-6 iterate until there are candidate splitters to be considered. One is selected and removed from `spls`, and the procedure Split is invoked to refine each block of $\mathcal{H}$ according to that splitter, and to generate new candidate ones.

**The `Split` procedure (Lines 1-7).** Each species $X$ has associated a real-valued field $X.\text{rr}$, initialized to 0 in Lines 2-3 in $O(s)$ time. Also, we assume that each product $\pi \in \Pi(R)$ is provided with a list, `inc`, which points to all the reactions that have that product. Each list `inc` has size $O(r)$, while exactly $r$ entries appear in all `inc`. The `inc` list allows to compute the reaction rates by iterating all reactions at most only once. In fact, given an input splitter $(\rho, \mathcal{M}_{spl})$, Lines 4-6 store $\mathbf{rr}[X + \rho, \mathcal{M}_{spl}]$ in $X.\text{rr}$, for each species $X$, by iterating once the `inc` list of each product multi-set $\pi \in \mathcal{M}_{spl}$. In particular, we can have two cases: either $\rho = \emptyset$, when only unary reactions are considered (as $X + \emptyset = X$), or $\rho = Z$, with $Z \in S$, when only binary reactions having $Z$ in their reagents are considered. In both cases, checking for the presence of $\rho$ in the reagents of each reaction takes constant time. The computation has $O(r)$ time complexity, since each reaction appears in $\pi.\text{inc}$ for one $\pi$ only.

**The actual splitting (Line 7).** Using the computed rates, Line 7 then performs the actual splitting. We do not detail this part, as it is inspired by the usual approach, e.g., [20, 3], consisting of the following three steps: (i) Each block is split using an associated balanced binary search tree (BST) in which each species $X$ of the block is inserted providing $\mathbf{rr}[X + \rho, \mathcal{M}_{spl}]$ as key, and a new block is added to $\mathcal{H}$ for each leaf of the BST; this requires $O(s \cdot \log s)$ time, as there are at most $s$ insertions in the BSTs, each having size at most $s$; (ii) If at least one block has been split, all candidate splitters must be discarded; this takes $O(r)$ time, as `spls` contains at most an entry per product $\pi \in \Pi(R)$, [2] while deletion from `spls` takes constant time assuming that it is implemented as a linked list; (iii) If at least a block has been split, all splitters have to be recomputed. This is because another multi-set lifting must be considered from the new partition. It takes $O(s \cdot r \cdot \log r)$ to do so. Thus, overall Split has time complexity $O(s \cdot (\log s + r \cdot \log r))$. Also, note that the BSTs do not worsen the space complexity, as only one for a block is built at a time.

---

[2] Recall that, given a block $\mathcal{M}_{spl}$, only one entry is stored to represent all pairs $(\rho, \mathcal{M}_{spl})$.

|       |        | *Original model* |        |          | *SMB reduction* |          |        |       |          |
|-------|--------|------------------|--------|----------|-----------------|----------|--------|-------|----------|
| Id    | Int.   | $|R|$            | $|S|$  | CTMC (s) | SMB (s)         | Red. (s) | $|R|$  | $|S|$ | CTMC (s) |
| M1    | 50     | 3538944          | 262146 | —        | 2.68E+5         | 2.04E+1  | 990    | 222   | 1.77E+1  |
| M2    | 50     | 786432           | 65538  | —        | 6.67E+3         | 4.09E+0  | 720    | 167   | 1.15E+1  |
| M3    | 50     | 172032           | 16386  | —        | 1.95E+2         | 3.4E–1   | 504    | 122   | 7.96E+0  |
| S1    | 50     | 36864            | 4098   | 9.12E+2  | 9.38E+0         | 1.01E–1  | 336    | 86    | 5.28E+0  |
| S2    | 50     | 7680             | 1026   | 1.03E+2  | 7.33E–1         | 3.00E–2  | 210    | 58    | 3.69E+0  |
| M5    | 3600   | 194054           | 14531  | 3.54E+5  | 5.88E+2         | 1.20E+0  | 142165 | 10855 | 3.29E+5  |
| M6    | 3840   | 187468           | 10734  | 1.79E+4  | 1.96E+2         | 5.68E–1  | 57508  | 3744  | 1.47E+3  |
| M7    | 3840   | 32776            | 2506   | 1.34E+3  | 8.80E+0         | 2.68E–1  | 16481  | 1281  | 4.88E+2  |
| S3    | 500000 | 284              | 143    | 4.13E+2  | 3.30E–2         | 1.40E–2  | 142    | 72    | 1.39E+2  |

**Table 1.** SMB reductions and corresponding speed-ups in CTMC simulation.

**Complexity.** We observe that Split is invoked $O(l \cdot s \cdot r)$ times. This is because, initially, $l \cdot r$ candidate splitters have to be considered. At every step where some blocks of $\mathcal{H}$ get split (which happens at most $s$ times), all splitters are removed, and at most $l \cdot r$ new candidate splitters are added to spls. In conclusion, syntactic Markovian bisimilarity takes $O(l \cdot s^2 \cdot r \cdot (\log s + r \cdot \log r))$ time and $O(s \cdot r)$ space.

*Remark 2.* We wish to stress that step (iii) of the actual splitting phase is not necessary in classic partition refinement algorithms [20, 3, 10]. This is because only blocks of the current partition are used as splitters in those algorithms. Hence, splitters are computed and maintained at no additional cost. It is exactly due to this reason that the time complexity of our algorithm exceeds those from [20, 3, 10].

## 6 Evaluation

In this section we experimentally evaluate SMB studying its effectiveness to reduce a number of biochemical models from the literature. All experiments were performed on a 2.6 GHz Intel Core i5 with 4 GB of RAM, and are replicable using a prototypal tool available at http://sysma.imtlucca.it/tools/erode/samba/. The tool takes in input CRNs specified in the .net format of BioNetGen [4], version 2.2.5-stable. The CRN reduced by SMB is then converted back to the BioNetGen format to perform stochastic simulations using Gillespie's stochastic simulation algorithm [28].

SMB was tested in terms of model reduction capabilities and corresponding CTMC analysis speed-ups on the collection of models listed in Table 1. Models with labels starting with "M" are the largest models also considered in [8]. Additionally, in this paper we analyze models S1–S3; M1–M3 and S1–S2 belong to the same family of synthetic benchmarks that are generated by varying the number of phosphorylation sites in a complex described in BioNetGen's rule-based format [40]. S3 arises by studying ultrasensitivity in multisite proteins [21]. In all cases we applied our reduction technique starting from the trivial partition with one block only (i.e., $\{S\}$ for every CRN $(S, R)$).

Column "*Int.*" shows the units of time used for the simulation, taken from the respective papers. This information was missing for M1–M3 and S1–S2, for which we used an estimate of their steady states. Initial populations for the simulations were taken as well from the respective papers. Under "*Original model*" are listed the number of reactions ($|R|$) and species ($|S|$) of the CRN, and the overall time to perform 10 simulations. The same information is given in the columns under "*SMB reduction*" for the corresponding CRNs reduced up to SMB, providing in addition the time necessary to compute the SMB partition (*SMB (s)*) and to perform the reduction (*Red. (s)*).

The results indicate that SMB can find equivalences in a significant number of models concerning different biological mechanisms. In the three largest models, M1–M3, SMB was able to provide a compact aggregated CRN which could be straightforwardly analyzed, while the simulations of the original models did not terminate due to out-of-memory errors in our experimental set-up. This is consistent with [40], where the same issue was reported for model M1. Models with more sensible reductions in the number of reactions gave better speed-ups. For example, for S1–S2 and M6–M7, the reduced CRN could be analyzed in about one tenth of the time necessary for the original one. We attribute this to the fact that at every simulation step, Gillespie's algorithm scans all reactions (in the worst case) to decide which one to fire next. Also, we note that typically many simulations, often in the order of hundreds or thousands, are necessary to satisfy a given confidence interval (or precision); hence, even small speed-ups per single run may turn into consistent gains in the overall simulation runtimes. Finally, as can be expected from their respective computational complexities, the runtimes to reduce a CRN by SMB according to Definition 7 are considerably smaller compared to the runtimes for computing the largest SMB.

**Comparison with $\kappa$-based reduction techniques.** SMB and stochastic fragmentation can be experimentally compared in rule-based biochemical models with finite underlying CRNs, like those in Table 1, where both techniques can be applied. In [8] we have shown that FB and *differential fragmentation*, a variant of fragmentation defined for the ODE semantics of $\kappa$, are not comparable. The same holds for SMB and stochastic fragmentation. For example, SMB reduces M12 from [8] to 56 species, while fragmentation does not. Conversely, the $\kappa$ model of cross-talk between a model of the early events of the EGF pathway and the insulin receptor of [16] can be reduced by stochastic fragmentation, but not by SMB.

**Comparison with Forward Bisimulation.** We now relate SMB with FB, introduced in [8] for the ODE semantics of CRNs. For this, it is convenient to recall such semantics. The ODE system $\dot{V} = F(V)$ underlying a CRN $(S, R)$ (where the dot notation indicates derivative with respect to time) is given by $F : \mathbb{R}_{\geq 0}^S \to \mathbb{R}^S$, where each component $F_X$, with $X \in S$ is defined by the expression

$$F_X(V) := \sum_{\rho \xrightarrow{\alpha} \pi \in R} (\pi(X) - \rho(X)) \cdot \alpha \cdot \prod_{Y \in S} V_Y^{\rho(Y)}. \tag{1}$$

This provides the well-known *mass-action* kinetics, where the reaction rate is proportional to the concentrations of the reactants involved.

*Example 7.* The ODE system associated to our running example is as follows.

$$\dot{V}_A = -8\,V_A \quad \dot{V}_C = 6\,V_A + 5\,V_C\,V_D + 6\,V_B + 10\,V_D\,V_E + 3\,V_D^2 \quad \dot{V}_E = -5\,V_D\,V_E$$
$$\dot{V}_B = -8\,V_B \quad \dot{V}_D = 6\,V_A + 6\,V_B - 6\,V_D^2$$

As for SMB, FB is an equivalence over the species of a CRN computed by looking at the reactions only. Also, FB induces a notion of lumpability similar in spirit to that of Definition 3, as it allows to rewrite the ODEs underlying a CRN in terms of macro-variables that govern the evolution of the cumulative concentrations of the species of each block.

*Example 8.* Consider the partition $\{\{A,B\},\{C,E\},\{D\}\}$ for our running example. This can be shown to be an FB. Indeed, the ODEs of $(S,R)$ can be rewritten, under the variable renaming $V_{AB} = V_A + V_B$, $V_{CE} = V_C + V_E$, as

$$\dot{V}_{AB} = -8\,V_{AB} \quad \dot{V}_{CE} = 6\,V_{AB} + 5\,V_{CE}\,V_D + 3\,V_D^2 \quad \dot{V}_D = 6\,V_{AB} - 6\,V_D^2$$

Although SMB and FB work on different semantics, the fact that they are both equivalences over species that induce analogous aggregations at the semantic level calls for the question of establishing a formal relation between the two equivalences.

**Theorem 3.** *Let $(S,R)$ be a CRN, and $\mathcal{R}$ an equivalence relation over $S$. Then, if $\mathcal{R}$ is an SMB for $(S,R)$, it also is an FB for $(S,R')$, with*

$$R' = \{X \xrightarrow{\alpha} \pi \mid (X \xrightarrow{\alpha} \pi) \in R\} \cup$$
$$\{X+Y \xrightarrow{\alpha} \pi \mid (X+Y \xrightarrow{\alpha} \pi) \in R \wedge X \neq Y\} \cup$$
$$\{X+X \xrightarrow{\alpha/2} \pi \mid (X+X \xrightarrow{\alpha} \pi) \in R\}$$

*When $R$ has singleton products only, then $\mathcal{R}$ is an SMB for $(S,R)$ iff it is an FB for $(S,R')$.*

An important remark to be made regarding this result is that it requires to halve the rates of homeoreactions. This is due to an inherent, well-known inconsistency existing between the CTMC and ODE semantics of CRNs. While, as discussed in Section 2, homeoreactions are treated specially in the CTMC semantics in order to capture the combinatorial nature of the discrete molecular interactions, the ODE semantics does not make such difference, e.g., [24], see also (1). We refer to [6] for a more in depth discussion on this. It is interesting to note that a *different* ODE semantics would be possible, grounded on a limit result by Kurtz which establishes the ODE solution as the asymptotic behavior of a sequence of infinitely large CTMCs induced by the same CRN with increasing volumes of a solution having given initial concentrations of species. This interpretation would lead to a $1/2$ coefficient in the rates of homeoreactions also in the ODE case [35]. We leave it for future work to understand if, by appropriately adapting FB to this different ODE semantics, Theorem 3 can be stated so as to relate SMB and FB *on the same CRN* also in presence of homeoreactions.

We also remark that the converse of the theorem does not hold in general for CRNs with non-singleton products: in our running example $\{\{A,B\},\{C,E\},\{D\}\}$ has been discussed to be an FB. However, given that $\mathbf{rr}(A + \emptyset, D) = 6$ and $\mathbf{rr}(B + \emptyset, D) = 0$, it is not an SMB.

FB has been applied to the models of Table 1 in [8], where a biological interpretation of the obtained aggregations was also provided. Interestingly, the largest SMBs of Table 1 correspond to the largest FBs of [8] on these models. Since none of these models have homeoreactions, we conclude that for these CRNs FB has the same discriminating power as SMB. Since S3 has unary products only, in that case this is guaranteed by Theorem 3.

## 7  Conclusion

Syntactic Markovian bisimulation (SMB) is an equivalence relation operating at the syntactic level of a chemical reaction network that induces a reduced one in the sense of the theory of Markov chain lumpability. A numerical evaluation has demonstrated its usefulness in practice by showing significant reductions in a number of models available in the literature, even if SMB is only a sufficient condition for aggregation. A partition-refinement algorithm computes the largest SMB that refines a given input partition. The freedom in choosing such input may be exploited to single out certain *observable species*. Thus, SMB may give a reduced model that exactly preserves the dynamics of interest. Since the CRN syntax is often combinatorially smaller than the underlying CTMC, we envisage SMB to be used as a pre-processing stage for CTMC analyses or for further reduction techniques on the semantics, either in exact or approximate form (e.g., [1]). Indeed, it would be interesting to conduct further experiments in order to understand how tight the lumping is with respect to the coarsest one obtained by applying Markov chain minimization algorithms on the fully enumerated state space; for this, we plan to develop fully integrated support for SMB into our software tool for model reduction techniques, *ERODE* [12] (http://sysma.imtlucca.it/tools/erode/).

We have shown that SMB is stricter than forward bisimulation (FB), a recently introduced bisimulation for the ODE semantics of CRNs. In a related research line, we developed differential bisimulation (DB) [34], a behavioural equivalence for the ODE semantics of a process algebra FEPA [44, 32]. In the future, we plan to provide a variant of DB which implies lumpability for the CTMC semantics of FEPA.

**Acknowledgments.** This work was partially supported by the EU project QUANTICOL, 600708. L. Cardelli is partially funded by a Royal Society Research Professorship.

## References

1. A. Abate, L. Brim, M. Ceska, and M.Z. Kwiatkowska. Adaptive aggregation of markov chains: Quantitative analysis of chemical reaction networks. In *CAV*, pages 195–213, 2015.
2. C. Autant and Ph. Schnoebelen. Place bisimulations in Petri nets. In *Application and Theory of Petri Nets*, 1992.
3. C. Baier, B. Engelen, and M. E. Majster-Cederbaum. Deciding bisimilarity and similarity for probabilistic processes. *J. Comput. Syst. Sci.*, 60(1):187–231, 2000.
4. M. L. Blinov, J. R. Faeder, B. Goldstein, and W. S. Hlavacek. BioNetGen: software for rule-based modeling of signal transduction based on the interactions of molecular domains. *Bioinformatics*, 20(17):3289–3291, 2004.

5. P. Buchholz. Exact and Ordinary Lumpability in Finite Markov Chains. *Journal of Applied Probability*, 31(1):59–75, 1994.
6. L. Cardelli. On process rate semantics. *Theor. Comput. Sci.*, 391(3):190–215, 2008.
7. L. Cardelli. Morphisms of reaction networks that couple structure to function. *BMC Systems Biology*, 8(1):84, 2014.
8. L. Cardelli, M. Tribastone, M. Tschaikowski, and A. Vandin. Forward and backward bisimulations for chemical reaction networks. In *CONCUR*, pages 226–239, 2015.
9. L. Cardelli, M. Tribastone, M. Tschaikowski, and A. Vandin. Comparing chemical reaction networks: A categorical and algorithmic perspective. In *LICS*, 2016. To appear.
10. L. Cardelli, M. Tribastone, M. Tschaikowski, and A. Vandin. Efficient syntax-driven lumping of differential equations. In *TACAS*, pages 93–111, 2016.
11. L. Cardelli, M. Tribastone, M. Tschaikowski, and A. Vandin. Symbolic computation of differential equivalences. In *POPL*, pages 137–150, 2016.
12. L. Cardelli, M. Tribastone, M. Tschaikowski, and A. Vandin. Erode: a tool for the evaluation and reduction of ordinary differential equations. In *TACAS*, 2017. To appear.
13. L. Cardelli and G. Zavattaro. Turing universality of the Biochemical Ground Form. *Mathematical Structures in Computer Science*, 20(1):45–73, 2010.
14. Luca Cardelli, Attila Csikász-Nagy, Neil Dalchau, Mirco Tribastone, and Max Tschaikowski. Noise reduction in complex biological switches. *Scientific Reports*, 6:20214, 2016.
15. F. Ciocchetta and J. Hillston. Bio-PEPA: A framework for the modelling and analysis of biological systems. *Theor. Comput. Sci.*, 410(33-34):3065–3084, 2009.
16. H. Conzelmann, D. Fey, and E. D. Gilles. Exact model reduction of combinatorial reaction networks. *BMC Systems Biology*, 2:78, 2008.
17. V. Danos, J. Feret, W. Fontana, R. Harmer, and J. Krivine. Abstracting the differential semantics of rule-based models: Exact and automated model reduction. In *LICS*, 2010.
18. V. Danos, J. Feret, W. Fontana, and J. Krivine. Scalable simulation of cellular signaling networks. In *APLAS*, 2007.
19. V. Danos and C. Laneve. Formal molecular biology. *TCS*, 325(1):69–110, 2004.
20. S. Derisavi, H. Hermanns, and W.H. Sanders. Optimal state-space lumping in Markov chains. *Inf. Process. Lett.*, 87(6):309–315, 2003.
21. O. Dushek, P.A. van der Merwe, and V. Shahrezaei. Ultrasensitivity in multisite phosphorylation of membrane-anchored proteins. *Biophysical Journal*, 100(5):1189–1197, 2011.
22. M. B. Elowitz, A. J. Levine, E. D. Siggia, and P. S. Swain. Stochastic gene expression in a single cell. *Science*, 297(5584):1183–1186, 2002.
23. J. R. Faeder, W. S. Hlavacek, I. Reischl, M. L. Blinov, H. Metzger, A. Redondo, C. Wofsy, and B. Goldstein. Investigation of early events in FcεRI-mediated signaling using a detailed mathematical model. *The Journal of Immunology*, 170(7):3769–3781, 2003.
24. M. Feinberg. Lectures on chemical reaction networks. Technical report, University of Wisconsin, 1979.
25. J. Feret, T. Henzinger, H. Koeppl, and T. Petrov. Lumpability abstractions of rule-based systems. *TCS*, 431:137–164, 2012.
26. S. Gay, S. Soliman, and F. Fages. A graphical method for reducing and relating models in systems biology. *Bioinformatics*, 26(18), 2010.
27. D. Gillespie. The chemical Langevin equation. *The Journal of Chemical Physics*, 113(1):297–306, 2000.
28. D.T. Gillespie. Exact stochastic simulation of coupled chemical reactions. *Journal of Physical Chemistry*, 81(25):2340–2361, December 1977.
29. S. Gilmore, J. Hillston, and M. Ribaudo. An efficient algorithm for aggregating PEPA models. *IEEE Transactions on Software Engineering*, 27(5):449–464, May 2001.
30. J. Heath, M. Kwiatkowska, G. Norman, D. Parker, and O. Tymchyshyn. Probabilistic model checking of complex biological pathways. *TCS*, 391(3):239–257, 2008.

31. H. Hermanns and M. Ribaudo. Exploiting symmetries in stochastic process algebras. In *European Simulation Multiconference*, pages 763–770, Manchester, UK, June 1998.
32. J. Hillston. *A Compositional Approach to Performance Modelling*. CUP, 1996.
33. D. T. Huynh and L. Tian. On some equivalence relations for probabilistic processes. *Fundam. Inform.*, 17(3):211–234, 1992.
34. Giulio Iacobelli, Mirco Tribastone, and Andrea Vandin. Differential bisimulation for a Markovian process algebra. In *MFCS*, pages 293–306, 2015.
35. T. G. Kurtz. The Relationship between Stochastic and Deterministic Models for Chemical Reactions. *Journal of Chemical Physics*, 57(7), 1972.
36. M. Kwiatkowska, G. Norman, and D. Parker. Symmetry reduction for probabilistic model checking. In *CAV*, pages 234–248, Berlin, Heidelberg, 2006. Springer-Verlag.
37. K. G. Larsen and A. Skou. Bisimulation through probabilistic testing. *Information and Computation*, 94(1):1–28, 1991.
38. R. Paige and R. Tarjan. Three partition refinement algorithms. *SIAM Journal on Computing*, 16(6):973–989, 1987.
39. Z. Rózsa and J. Tóth. Exact linear lumping in abstract spaces. *Electronic Journal of Qualitative Theory of Differential Equations [electronic only]*, 2003.
40. M. W. Sneddon, J. R. Faeder, and T. Emonet. Efficient modeling, simulation and coarse-graining of biological complexity with NFsim. *Nature Methods*, 8(2):177–183, 2011.
41. D. Soloveichik, M. Cook, E. Winfree, and J. Bruck. Computation with finite stochastic chemical reaction networks. *Natural Computing*, 7(4):615–633, 2008.
42. R. Suderman and E. J. Deeds. Machines vs. ensembles: Effective MAPK signaling through heterogeneous sets of protein complexes. *PLoS Comput Biol*, 9(10):e1003278, 10 2013.
43. S. Tognazzi, M. Tribastone, M. Tschaikowski, and A. Vandin. EGAC: A Genetic Algorithm to Compare Chemical Reaction Networks. In *The Genetic and Evolutionary Computation Conference (GECCO)*, 2017. To appear.
44. M. Tschaikowski and M. Tribastone. Exact fluid lumpability for Markovian process algebra. In *23rd International Conference on Concurrency Theory (CONCUR)*, LNCS, pages 380–394, 2012.
45. M. Tschaikowski and M. Tribastone. Generalised Communication for Interacting Agents. In *QEST*, pages 178–188, September 2012.
46. Max Tschaikowski and Mirco Tribastone. Approximate Reduction of Heterogenous Nonlinear Models With Differential Hulls. *IEEE Trans. Automat. Contr.*, 61(4):1099–1104, 2016.
47. Andrea Vandin and Mirco Tribastone. Quantitative abstractions for collective adaptive systems. In *SFM 2016, Bertinoro Summer School*, pages 202–232, 2016.
48. G. Zavattaro and L. Cardelli. Termination problems in chemical kinetics. In *CONCUR*, pages 477–491, 2008.

## A  SMB properties

This section collects the results regarding syntactic Markovian bisimilariy, and its relation with forward bisimulation.

**Proposition 1 (Reduced CRNs induce lumped CTMCs).** *Let $(S, R)$ be a CRN, $I$ a set of indices, and $\mathcal{R}_i$ an SMB for $(S, R)$, for all $i \in I$. The transitive closure of their union $\mathcal{R} = (\bigcup_{i \in I} \mathcal{R}_i)^*$ is an SMB for $(S, R)$.*

*Proof.* We first note that $\mathcal{R}$ is an equivalence relation over $S$, as it is the transitive closure of the union of equivalence relations over $S$. For $i \in I$, let $\mathcal{H}_i$ denote the partition induced over $S$ by $\mathcal{R}_i$, and $\mathcal{H}$ the one induced by $\mathcal{R}$. For any $i \in I$, any block $H^i \in \mathcal{H}_i$ is contained in a block $H \in \mathcal{H}$, implying that any $H \in \mathcal{H}$ is the union of blocks of $\mathcal{H}_i$. For $(X_1, X_2) \in \mathcal{R}$, we have that $(X_1, X_2) \in (\bigcup_{i \in I} \mathcal{R}_i)^n$, for some $n > 0$. We now show that $\mathcal{R}$ is an SMB by induction over $n$. Let $\mathcal{R}^n$ be $(\bigcup_{i \in I} \mathcal{R}_i)^n$, and $\rho \in \mathcal{MS}(S)$.

*Base case ($n = 1$):* $(X_1, X_2) \in \mathcal{R}^1$ implies that $(X_1, X_2) \in \mathcal{R}_i$, for some $i \in I$. In order to prove that the condition required by SMB in Definition 6 holds, we use that for any $H \in \mathcal{H}$ and any $i \in I$ we have that there exists some set of indices $J^i$ such that $H = \bigcup_{j \in J^i} H^i_j$, with $H^i_j$ a block of $\mathcal{H}_i$; hence, $\mathbf{rr}[X_1 + \rho, H] = \sum_{j \in J^i} \mathbf{rr}[X_1 + \rho, H^i_j]$.

*Inductive step:* we assume that the condition required by SMB holds for $\mathcal{R}^m$, $\forall m < n$. If $(X_1, X_2) \in \mathcal{R}^n$, then there exists an $X_3 \in S$ such that $(X_1, X_3) \in \mathcal{R}_i$ for some $i \in I$, and $(X_3, X_2) \in \mathcal{R}^{n-1}$. Then, the claim follows from a similar argument as in the base case and the induction hypothesis (for more details see, e.g., Proposition 8.2.1 of [32] ).

Before proving the proof of Theorem 3 we recall the notion of forward bisimulation from [8].

**Definition 8 (Reaction and production rates [8]).** *Let $(S, R)$ be a CRN, $X, Y \in S$, and $\rho \in \mathcal{MS}(S)$. The $\rho$-reaction rate of $X$, and the $\rho$-production rate of $Y$-elements by $X$ are defined respectively as*

$$\mathbf{ccr}[X, \rho] := (\rho(X) + 1) \sum_{X+\rho \xrightarrow{\alpha} \pi \in R} \alpha, \quad \mathbf{pr}(X, \rho, Y) := (\rho(X) + 1) \sum_{X+\rho \xrightarrow{\alpha} \pi \in R} \alpha \cdot \pi(Y)$$

*Finally, for $H \subseteq S$ we define $\mathbf{pr}[X, \rho, H] := \sum_{Y \in H} \mathbf{pr}(X, \rho, Y)$.*

**Definition 9 (Forward CRN Bisimulation [8]).** *Let $(S, R)$ be a CRN, $\mathcal{R}$ an equivalence relation over $S$ and $\mathcal{H} = S/\mathcal{R}$. Then, $\mathcal{R}$ is a forward CRN bisimulation (abbreviated FB) if for all $(X, Y) \in \mathcal{R}$, all $\rho \in \mathcal{MS}(S)$, and all $H \in \mathcal{H}$ it holds that*

$$\mathbf{ccr}[X, \rho] = \mathbf{ccr}[Y, \rho] \quad \textit{and} \quad \mathbf{pr}[X, \rho, H] = \mathbf{pr}[Y, \rho, H] \tag{2}$$

**Theorem 3 (SMB discriminates more than FB).** *Let $(S, R)$ be a CRN, and $\mathcal{R}$ an equivalence relation over $S$. Then, if $\mathcal{R}$ is an SMB for $(S, R)$, it also is an FB for $(S, R')$,*

*with*

$$R' = \{X \xrightarrow{\alpha} \pi \mid (X \xrightarrow{\alpha} \pi) \in R\} \cup$$
$$\{X + Y \xrightarrow{\alpha} \pi \mid (X + Y \xrightarrow{\alpha} \pi) \in R \land X \neq Y\} \cup$$
$$\{X + X \xrightarrow{\alpha/2} \pi \mid (X + X \xrightarrow{\alpha} \pi) \in R\}$$

*When $R$ has singleton products only, then $\mathcal{R}$ is an SMB for $(S, R)$ iff it is an FB for $(S, R')$.*

*Proof.* Note that the factor $\rho(X) + 1$ in **ccr** and **pr** has the effect of doubling the rates of homeoreactions in $R'$ only, obtaining back their original rates in $R$. It hence easy to see that $\mathbf{ccr}_{(S,R')}[X, \rho] = \mathbf{ccr}'_{(S,R)}[X, \rho]$ and $\mathbf{pr}_{(S,R')}(X, \rho, Y) = \mathbf{pr}'_{(S,R)}(X, \rho, Y)$, with

$$\mathbf{ccr}'[X, \rho] := \sum_{X+\rho \xrightarrow{\alpha} \pi \in R} \alpha, \qquad \mathbf{pr}'(X, \rho, Y) := \sum_{X+\rho \xrightarrow{\alpha} \pi \in R} \alpha \cdot \pi(Y)$$

In the rest of the proof we will hence use $\mathbf{ccr}'_{(S,R)}$ and $\mathbf{pr}'_{(S,R)}$, rather than $\mathbf{ccr}_{(S,R')}$ and $\mathbf{pr}_{(S,R')}$, respectively. Since all **rr**, **ccr**' and **pr**' are defined over $(S, R)$, we avoid to write the CRN as prefix.

Let $\mathcal{H}$ be the partition induced by $\mathcal{R}$ over $S$, and $\mathcal{H}^\uparrow$ the partition induced by $\mathcal{R}^\uparrow$ over $\mathcal{MS}(S)$. In order to close the proof we show that the condition of SMB in $(S, R)$ implies both conditions (i) and (ii) of FB in $(S, R)$. We start with condition (i) of FB. Let $(X, Y) \in \mathcal{R}$.

For each $\rho \in \mathcal{MS}(S)$, what we want to show is that $\mathbf{rr}[X + \rho, \mathcal{M}] = \mathbf{rr}[Y + \rho, \mathcal{M}]$ $\forall \mathcal{M} \in \mathcal{H}^\uparrow$, implies $\mathbf{ccr}'[X, \rho] = \mathbf{ccr}'[Y, \rho]$. This trivially holds because

$$\sum_{\mathcal{M} \in \mathcal{MS}(S)} \mathbf{rr}[X + \rho, \mathcal{M}] = \mathbf{ccr}'[X, \rho]$$

We now focus on condition (ii). What we want to show is that $\mathbf{rr}[X + \rho, \mathcal{M}] = \mathbf{rr}[Y + \rho, \mathcal{M}]$ $\forall \mathcal{M} \in \mathcal{H}^\uparrow$, implies $\mathbf{pr}'[X, \rho, H] = \mathbf{pr}'[Y, \rho, H]$ $\forall H \in \mathcal{H}$. All multi-sets in $\mathcal{M}$ have same number of elements of species of each block $H \in \mathcal{H}$. Thus, with a slight abuse of notation we can write $\mathcal{M}(H)$ to denote $\sum_{Z \in H} \pi^{\mathcal{M}}(Z)$, with $\pi^{\mathcal{M}}$ any multi-set in $\mathcal{M}$. Then, a reaction with reagents $X + \rho$, rate $r$ and product any $\pi \in \mathcal{M}$, produces elements of species in $H$ with rate $\mathcal{M}(H) \cdot r$. Finally, for each $H \in \mathcal{H}$ we easily obtain $\mathbf{pr}'[X, \rho, H] = \sum_{\mathcal{M} \in \mathcal{H}^\uparrow} \mathcal{M}(H) \cdot \mathbf{rr}[X + \rho, \mathcal{M}]$, and similarly for $Y$, allowing us to conclude $\mathbf{pr}'[X, \rho, H] = \mathbf{pr}'[Y, \rho, H]$, closing the proof.

The "only if" direction for the case of CRNs with unary products follows from the fact that condition (ii) of FB implies condition (i), while $\mathbf{pr}'$ and $\mathbf{rr}$ degenerate to the same notion.

## B  SMB and ordinary CTMC Lumpability (Theorem 1)

We hereby provide the technical results relating SMB to ordinary CTMC lumpability. We start providing a proposition and a remark used in the proof of Theorem 1. For $(S, R)$ a CRN, $H \subseteq S$ and $\sigma \in \mathcal{MS}(S)$, we shall use $\sigma(H)$ for $\sum_{X \in H} \sigma(X)$.

**Proposition 2.** *Let $(S, R)$ be a CRN, $\mathcal{R}$ an SMB, $\mathcal{R}^\uparrow$ its multi-set lifting, and $\mathcal{H} = S/\mathcal{R}$. For any $(\sigma, \sigma') \in \mathcal{R}^\uparrow$ and any $H, \tilde{H} \in \mathcal{H}$ it holds*

$$\sum_{\substack{X \in H \\ \sigma(X) > 0}} \sigma(X) \sum_{\substack{Y \in \tilde{H} \\ (\sigma - X)(Y) > 0}} (\sigma - X)(Y) = \sum_{\substack{X \in H \\ \sigma'(X) > 0}} \sigma'(X) \sum_{\substack{Y \in \tilde{H} \\ (\sigma' - X)(Y) > 0}} (\sigma' - X)(Y) \quad (3)$$

*Proof.* We distinguish among two cases: $H \neq \tilde{H}$ and $H = \tilde{H}$. For the case $H \neq \tilde{H}$ we can rewrite the left-hand-side of Equation (3) as:

$$\sum_{X \in H \text{ s.t. } \sigma(X) > 0} \sigma(X) \sum_{Y \in \tilde{H} \text{ s.t. } \sigma(Y) > 0} \sigma(Y) = \sigma(H) \cdot \sigma(\tilde{H})$$

The same holds for $\sigma'$, allowing to rewrite Equation (3) (for the case $H \neq \tilde{H}$) as $\sigma(H) \cdot \sigma(\tilde{H}) = \sigma'(H) \cdot \sigma'(\tilde{H})$, which directly follows from the fact that $(\sigma, \sigma') \in \mathcal{R}^\uparrow$.

As regards the case $H = \tilde{H}$, we can rewrite the left-hand-side of Equation (3) as

$$\sum_{X \in H \text{ s.t. } \sigma(X) > 0} \sigma(X) \sum_{Y \in H \text{ s.t. } (\sigma - X)(Y) > 0} (\sigma - X)(Y), \text{ which in turn can be rewritten as}$$

$$\sum_{X \in H \text{ s.t. } \sigma(X) > 0} \sigma(X) \cdot (\sigma - X)(H) = \sum_{X \in H \text{ s.t. } \sigma(X) > 0} \sigma(X) \cdot (\sigma(H) - 1) = \sigma(H) \cdot (\sigma(H) - 1)$$

The same holds for $\sigma'$, allowing to rewrite Equation (3) (for the case $H = \tilde{H}$) as $\sigma(H) \cdot (\sigma(H) - 1) = \sigma'(H) \cdot (\sigma'(H) - 1)$, which follows from $(\sigma, \sigma') \in \mathcal{R}^\uparrow$. □

**Fact 4** *Let $(S, R)$ be a CRN and $\mathcal{R}$ an SMB. Let $\mathcal{H}$ and $\mathcal{H}^\uparrow$ be the partitions induced by $\mathcal{R}$ on $S$, and by $\mathcal{R}^\uparrow$ on $\mathcal{MS}(S)$, respectively. For all $\sigma, \sigma_2, \pi, \pi_2 \in \mathcal{MS}(S)$, we have*

- $(\sigma \cup \pi, \sigma \cup \pi_2) \in \mathcal{R}^\uparrow$ *if and only if* $(\pi, \pi_2) \in \mathcal{R}^\uparrow$,
- *if* $(\sigma, \sigma_2) \in \mathcal{R}^\uparrow$, *then* $(\sigma \cup \pi, \sigma_2 \cup \pi_2) \in \mathcal{R}^\uparrow$ *if and only if* $(\pi, \pi_2) \in \mathcal{R}^\uparrow$.

*Thus, for any $\mathcal{M}, \tilde{\mathcal{M}} \in \mathcal{H}^\uparrow$, if it is possible to obtain multi-sets in $\tilde{\mathcal{M}}$ by adding species to those in $\mathcal{M}$, i.e., if $\mathcal{M}(H) \leq \tilde{\mathcal{M}}(H)$ for all $H \in \mathcal{H}$ (where $\mathcal{M}(H)$ is the equal number of $H$-elements of each multi-set in $\mathcal{M}$), then there exists exactly one $\hat{\mathcal{M}} \in \mathcal{H}^\uparrow$ such that by pairwise merging each $\sigma \in \mathcal{M}$ with each $\hat{\sigma} \in \hat{\mathcal{M}}$ we obtain all $\tilde{\sigma} \in \tilde{\mathcal{M}}$.*
□

These intermediate results allow us to provide the proof of Theorem 1.

**Theorem 1.** *Let $\mathcal{R}$ be an SMB for the CRN $(S, R)$. Then, its multi-set lifting $\mathcal{R}^\uparrow$ induces the ordinarily lumpable partition $\mathcal{H}^\uparrow$ on $MC(\sigma_0)$ for any initial state $\sigma_0$.*

*Proof.* We have to prove that for any $\mathcal{M}, \tilde{\mathcal{M}} \in \mathcal{H}^\uparrow$ and $\sigma, \sigma' \in \mathcal{M}$ we have $q[\sigma, \tilde{\mathcal{M}}] = q[\sigma', \tilde{\mathcal{M}}]$. We may have either $\mathcal{M} \neq \tilde{\mathcal{M}}$, or $\mathcal{M} = \tilde{\mathcal{M}}$. We start with the $\mathcal{M} \neq \tilde{\mathcal{M}}$ case.
By Definitions 1, 2 we have

$$q[\sigma, \tilde{\mathcal{M}}] = \sum_{\theta \in \tilde{\mathcal{M}}} q(\sigma, \theta) = \sum_{\theta \in \tilde{\mathcal{M}}} \sum_{\sigma \xrightarrow{r} \theta \in MTS(\sigma_0)} r = \sum_{\theta \in \tilde{\mathcal{M}}} \sum_{\sigma \xrightarrow{r} \theta \in out(\sigma)} r \,,$$

which in turn is equal to

$$\sum_{\theta \in \tilde{\mathcal{M}}} \sum_{\substack{X \xrightarrow{r} \pi \in R \\ (\sigma-X)+\pi=\theta}} \sigma(X) \cdot r + \sum_{\theta \in \tilde{\mathcal{M}}} \sum_{\substack{X+Y \xrightarrow{r} \pi \in R \\ (\sigma-(X+Y))+\pi=\theta}} \sigma(X) \cdot (\sigma-X)(Y) \cdot r =$$

$$\sum_{\substack{X \xrightarrow{r} \pi \in R \\ (\sigma-X)+\pi \in \tilde{\mathcal{M}}}} \sigma(X) \cdot r + \tag{4}$$

$$\sum_{\substack{X+Y \xrightarrow{r} \pi \in R, X \neq Y \\ (\sigma-(X+Y))+\pi \in \tilde{\mathcal{M}}}} \sigma(X) \cdot (\sigma-X)(Y) \cdot r + \tag{5}$$

$$\sum_{\substack{X+X \xrightarrow{r} \pi \in R \\ (\sigma-(X+X))+\pi \in \tilde{\mathcal{M}}}} \sigma(X) \cdot (\sigma-X)(X) \cdot \frac{r}{2} \tag{6}$$

We first focus on Equation (4), which can be rewritten as

$$\sum_{H \in \mathcal{H}} \sum_{X \in H} \sigma(X) \sum_{\substack{X \xrightarrow{r} \pi \in R \\ (\sigma-X)+\pi \in \tilde{\mathcal{M}}}} r = \sum_{H \in \mathcal{H}} \sum_{X \in H \text{ s.t. } \sigma(X)>0} \sigma(X) \sum_{\substack{X \xrightarrow{r} \pi \in R \\ (\sigma-X)+\pi \in \tilde{\mathcal{M}}}} r \tag{7}$$

Now, for all $H \in \mathcal{H}$, for all $X, X' \in H$ with $\sigma(X) > 0$ and $\sigma(X') > 0$, we have $(\sigma - X, \sigma - X') \in \mathcal{R}^{\uparrow}$. Thus, by Fact 4 [3] we have that there exists a $\hat{\mathcal{M}}_H \in \mathcal{H}^{\uparrow}$ such that, for each $X \in H$ with $\sigma(X) > 0$, merging $\sigma - X$ with each $\pi \in \hat{\mathcal{M}}_H$ we obtain all the multi-sets in $\tilde{\mathcal{M}}$ considered in the right-hand-side of Equation (7), which can thus be rewritten as

$$\sum_{H \in \mathcal{H}} \sum_{X \in H \text{ s.t. } \sigma(X)>0} \sigma(X) \sum_{\pi \in \hat{\mathcal{M}}_H} \sum_{X \xrightarrow{r} \pi \in R} r =$$

$$\sum_{H \in \mathcal{H}} \sum_{X \in H \text{ s.t. } \sigma(X)>0} \sigma(X) \cdot \mathbf{rr}[X + \emptyset, \hat{\mathcal{M}}_H] =$$

$$\sum_{H \in \mathcal{H}} \sum_{X \in H} \sigma(X) \cdot \mathbf{rr}[X + \emptyset, \hat{\mathcal{M}}_H] = \quad \text{by Definition 6}$$

$$\sum_{H \in \mathcal{H}} \mathbf{rr}[X^H + \emptyset, \hat{\mathcal{M}}_H] \sum_{X \in H} \sigma(X) = \sum_{H \in \mathcal{H}} \mathbf{rr}[X^H + \emptyset, \hat{\mathcal{M}}_H] \cdot \sigma(H)$$

The same holds for $\sigma'$, obtaining

$$\sum_{X \xrightarrow{r} \pi \in R \text{ s.t. } (\sigma'-X)+\pi \in \tilde{\mathcal{M}}} \sigma'(X) \cdot r = \sum_{H \in \mathcal{H}} \mathbf{rr}[X^H + \emptyset, \hat{\mathcal{M}}_H] \cdot \sigma'(H)$$

---

[3] We can apply the remark only if $\mathcal{M}(H') \leq \tilde{\mathcal{M}}(H')$ for all $H' \neq H$, and $\mathcal{M}(H)-1 \leq \tilde{\mathcal{M}}(H)$. In the cases in which this does not hold, we have $q[\sigma, \tilde{\mathcal{M}}] = q[\sigma', \tilde{\mathcal{M}}] = 0$, closing the proof.

For all $H \in \mathcal{H}$, for all $X, X' \in H$ with $\sigma(X) > 0$ and $\sigma'(X') > 0$ we have $(\sigma - X, \sigma' - X') \in \mathcal{R}^\uparrow$, implying that the considered $\hat{\mathcal{M}}_H$ are the same for $\sigma$ and $\sigma'$. This allows us to close the case, as, for any $H \in \mathcal{H}$, $\mathbf{rr}[X^H + \emptyset, \hat{\mathcal{M}}_H]$ does not depend on $\sigma$ or $\sigma'$, while by $(\sigma, \sigma') \in \mathcal{R}^\uparrow$, by Definition 4 we have $\sigma(H) = \sigma'(H)$ for all $H \in \mathcal{H}$.

We now address Equations (5), (6), i.e., we show that

$$\sum_{\substack{X+Y \xrightarrow{r} \pi \in R, X \neq Y \\ (\sigma - (X+Y)) + \pi \in \tilde{\mathcal{M}}}} \sigma(X) \cdot (\sigma - X)(Y) \cdot r + \sum_{\substack{X+X \xrightarrow{r} \pi \in R \\ (\sigma - (X+X)) + \pi \in \tilde{\mathcal{M}}}} \sigma(X) \cdot (\sigma - X)(X) \cdot \frac{r}{2} = \tag{8}$$

$$\sum_{\substack{X+Y \xrightarrow{r} \pi \in R, X \neq Y \\ (\sigma' - (X+Y)) + \pi \in \tilde{\mathcal{M}}}} \sigma'(X) \cdot (\sigma' - X)(Y) \cdot r + \sum_{\substack{X+X \xrightarrow{r} \pi \in R \\ (\sigma' - (X+X)) + \pi \in \tilde{\mathcal{M}}}} \sigma'(X) \cdot (\sigma' - X)(X) \cdot \frac{r}{2} \tag{9}$$

We start from the second summands. We can rewrite the second summand of Equation (8) as:

$$\sum_{H \in \mathcal{H}} \sum_{X \in H \text{ s.t. } \sigma(X) > 1} \sigma(X) \cdot (\sigma - X)(X) \sum_{\substack{X+X \xrightarrow{r} \pi \in R \\ (\sigma - (X+X)) + \pi \in \tilde{\mathcal{M}}}} \frac{r}{2} \tag{10}$$

Similarly to what said for Equation (7), for all $H \in \mathcal{H}$, for all $X, X' \in H$ with $\sigma(X) > 1$ and $\sigma(X') > 1$ we have $(\sigma - (X+X), \sigma - (X'+X')) \in \mathcal{R}^\uparrow$. Thus, by Fact 4 we have that there exists a block $\hat{\mathcal{M}}_{HH} \in \mathcal{H}^\uparrow$ such that, for each $X \in H$ with $\sigma(X) > 1$, merging $\sigma - (X+X)$ with each $\pi \in \hat{\mathcal{M}}_{HH}$ we obtain all the multi-sets in $\tilde{\mathcal{M}}$ considered in Equation (10), which can thus be rewritten as

$$\frac{1}{2} \sum_{H \in \mathcal{H}} \sum_{X \in H \text{ s.t. } \sigma(X) > 1} \sigma(X) \cdot (\sigma - X)(X) \cdot \sum_{\pi \in \hat{\mathcal{M}}_{HH}} \sum_{X+X \xrightarrow{r} \pi \in R} r =$$

$$\frac{1}{2} \sum_{H \in \mathcal{H}} \sum_{X \in H \text{ s.t. } \sigma(X) > 1} \sigma(X) \cdot (\sigma - X)(X) \cdot \mathbf{rr}[X+X, \hat{\mathcal{M}}_{HH}] \tag{11}$$

As regards the first summand of Equation (8), we can rewrite it as follows, where $1/2$ compensates the fact that each $X + Y$ is considered twice (e.g., $X \in H, Y \in \tilde{H}$, and $Y \in H, X \in \tilde{H}$):

$$\sum_{\substack{X+Y \xrightarrow{r} \pi \in R \text{ s.t. } X \neq Y \\ (\sigma - (X+Y)) + \pi \in \tilde{\mathcal{M}}}} \sigma(X) \cdot \sigma(Y) \cdot r = \frac{1}{2} \sum_{H \in \mathcal{H}} \sum_{\substack{X \in H \\ \sigma(X) > 0}} \sigma(X) \sum_{\tilde{H} \in \mathcal{H}} \sum_{\substack{Y \in \tilde{H}, Y \neq X, \\ \sigma(Y) > 0}} \sigma(Y) \sum_{\substack{X+Y \xrightarrow{r} \pi \in R \\ (\sigma - (X+Y)) + \pi \in \tilde{\mathcal{M}}}} r \tag{12}$$

Similarly to what said for Equation (7) and Equation (10), for all $H \in \mathcal{H}$, for all $X, X' \in H$ with $\sigma(X) > 0$ and $\sigma(X') > 0$, as well as for all $\tilde{H} \in \mathcal{H}$, for all $Y, Y' \in \tilde{H}$ with $\sigma(Y) > 0$, $\sigma(Y') > 0$, $X \neq Y$ and $X' \neq Y'$, we have $((\sigma - X) - Y, (\sigma - X') - Y') \in \mathcal{R}^\uparrow$. Thus, by Fact 4 we have that there exists a block $\hat{\mathcal{M}}_{H\tilde{H}} \in \mathcal{H}^\uparrow$ such that, for each $X \in H$ with $\sigma(X) > 0$, for each $Y \in \tilde{H}$ with $X \neq Y$ and $\sigma(Y) > 0$, merging $(\sigma - X) - Y$ with each $\pi \in \hat{\mathcal{M}}_{H\tilde{H}}$ we obtain all the multi-sets in $\hat{\mathcal{M}}$ considered in the right-hand side of Equation (12), which can thus be rewritten as

$$\frac{1}{2} \sum_{H \in \mathcal{H}} \sum_{\substack{X \in H \\ \sigma(X) > 0}} \sigma(X) \sum_{\tilde{H} \in \mathcal{H}} \sum_{\substack{Y \in \tilde{H}, Y \neq X, \\ \sigma(Y) > 0}} \sigma(Y) \sum_{\pi \in \hat{\mathcal{M}}_{H\tilde{H}}} \sum_{X+Y \xrightarrow{r} \pi \in R} r =$$

$$\frac{1}{2} \sum_{H \in \mathcal{H}} \sum_{\substack{X \in H \\ \sigma(X) > 0}} \sigma(X) \sum_{\tilde{H} \in \mathcal{H}} \sum_{\substack{Y \in \tilde{H}, Y \neq X, \\ \sigma(Y) > 0}} \sigma(Y) \cdot \mathbf{rr}[X + Y, \hat{\mathcal{M}}_{H\tilde{H}}] \quad (13)$$

By putting back together the case $X = Y$ (Equation (11)), and the case $X \neq Y$ (Equation (13)), we obtain

$$\frac{1}{2} \sum_{H \in \mathcal{H}} \sum_{\substack{X \in H \\ \sigma(X) > 0}} \sigma(X) \sum_{\tilde{H} \in \mathcal{H}} \sum_{\substack{Y \in \tilde{H}, Y \neq X, \\ \sigma(Y) > 0}} \sigma(Y) \cdot \mathbf{rr}[X + Y, \hat{\mathcal{M}}_{H\tilde{H}}] +$$

$$\frac{1}{2} \sum_{H \in \mathcal{H}} \sum_{X \in H \text{ s.t. } \sigma(X) > 1} \sigma(X) \cdot (\sigma - X)(X) \cdot \mathbf{rr}[X + X, \hat{\mathcal{M}}_{HH}] =$$

$$\frac{1}{2} \sum_{H \in \mathcal{H}} \sum_{\substack{X \in H \\ \sigma(X) > 0}} \sigma(X) \sum_{\tilde{H} \in \mathcal{H}} \sum_{\substack{Y \in \tilde{H} \\ (\sigma - X)(Y) > 0}} (\sigma - X)(Y) \cdot \mathbf{rr}[X + Y, \hat{\mathcal{M}}_{H\tilde{H}}] = \text{ (by Definition 6 on } Y\text{)}$$

$$\frac{1}{2} \sum_{H \in \mathcal{H}} \sum_{\substack{X \in H \\ \sigma(X) > 0}} \sigma(X) \sum_{\tilde{H} \in \mathcal{H}} \mathbf{rr}[Y^{\tilde{H}} + X, \hat{\mathcal{M}}_{H\tilde{H}}] \sum_{\substack{Y \in \tilde{H} \\ (\sigma - X)(Y) > 0}} (\sigma - X)(Y) =$$

$$\frac{1}{2} \sum_{H \in \mathcal{H}} \sum_{\tilde{H} \in \mathcal{H}} \sum_{\substack{X \in H \\ \sigma(X) > 0}} \sigma(X) \cdot \mathbf{rr}[Y^{\tilde{H}} + X, \hat{\mathcal{M}}_{H\tilde{H}}] \sum_{\substack{Y \in \tilde{H} \\ (\sigma - X)(Y) > 0}} (\sigma - X)(Y) = \text{ (by Definition 6 on } X\text{)}$$

$$\frac{1}{2} \sum_{H \in \mathcal{H}} \sum_{\tilde{H} \in \mathcal{H}} \mathbf{rr}[X^H + Y^{\tilde{H}}, \hat{\mathcal{M}}_{H\tilde{H}}] \sum_{\substack{X \in H \\ \sigma(X) > 0}} \sigma(X) \sum_{\substack{Y \in \tilde{H} \\ (\sigma - X)(Y) > 0}} (\sigma - X)(Y) \quad (14)$$

We have thus rewritten Equation (8) as as Equation (14). The same holds for $\sigma'$, allowing to rewrite Equation (9) as

$$\frac{1}{2} \sum_{H \in \mathcal{H}} \sum_{\tilde{H} \in \mathcal{H}} \mathbf{rr}[X^H + Y^{\tilde{H}}, \hat{\mathcal{M}}_{H\tilde{H}}] \sum_{\substack{X \in H \\ \sigma'(X) > 0}} \sigma'(X) \sum_{\substack{Y \in \tilde{H} \\ (\sigma' - X)(Y) > 0}} (\sigma' - X)(Y) \quad (15)$$

This allows to close the case, as the $\mathbf{rr}$ in Equations (14), (15) do not depend on $\sigma$ or $\sigma'$, and for any $(\sigma, \sigma') \in \mathcal{R}^\uparrow$, by Proposition 2 we have that for any $H, \tilde{H} \in \mathcal{H}$:

$$\sum_{X \in H \text{ s.t. } \sigma(X) > 0} \sigma(X) \sum_{Y \in \tilde{H} \text{ s.t. } (\sigma - X)(Y) > 0} (\sigma - X)(Y) = \sum_{X \in H \text{ s.t. } \sigma'(X) > 0} \sigma'(X) \sum_{Y \in \tilde{H} \text{ s.t. } (\sigma' - X)(Y) > 0} (\sigma' - X)(Y)$$

Note that for all $H \in \mathcal{H}$, for all $X, X' \in H$ with $\sigma(X) > 0$ and $\sigma'(X') > 0$, as well as for all $\tilde{H} \in \mathcal{H}$, for all $Y, Y' \in \tilde{H}$ with $(\sigma - X)(Y) > 0$, and $(\sigma' - X')(Y') > 0$ we have $((\sigma - X) - Y, (\sigma' - X') - Y') \in \mathcal{R}^\uparrow$, implying that the considered $\mathcal{M}_{H\tilde{H}}$ are the same for $\sigma$ and $\sigma'$.

We have closed the case $\mathcal{M} \neq \tilde{\mathcal{M}}$. We now address the case $\mathcal{M} = \tilde{\mathcal{M}}$. We have to prove that $q[\sigma, \mathcal{M}] = q[\sigma', \mathcal{M}]$, which, given that $\sigma, \sigma' \in \mathcal{M}$, can be rewritten as

$$q[\sigma, \mathcal{M} \setminus \{\sigma\}] + q(\sigma, \sigma) = q[\sigma', \mathcal{M} \setminus \{\sigma'\}] + q(\sigma', \sigma') . \tag{16}$$

From Definition 2 we have $q(\sigma, \sigma) = -q[\sigma, \mathcal{MS}(S) \setminus \{\sigma\}]$. If we partition $\mathcal{MS}(\mathcal{M})$ according to $\mathcal{H}^\uparrow$, we obtain $q(\sigma, \sigma) = -q[\sigma, \mathcal{M} \setminus \{\sigma\}] - \sum_{\widehat{\mathcal{M}} \in \mathcal{H}^\uparrow} q[\sigma, \widehat{\mathcal{M}}]$

The same holds for $\sigma'$, allowing us to rewrite Equation (16) as

$$- \sum_{\widehat{\mathcal{M}} \in \mathcal{H}^\uparrow} q[\sigma, \widehat{\mathcal{M}}] = - \sum_{\widehat{\mathcal{M}} \in \mathcal{H}^\uparrow} q[\sigma', \widehat{\mathcal{M}}] \tag{17}$$

Finally, we close the proof noticing that Equation (17) follows from the case $\mathcal{M} \neq \tilde{\mathcal{M}}$. In fact, we have shown that for every $\widehat{\mathcal{M}} \neq \mathcal{M}$, we have $q[\sigma, \widehat{\mathcal{M}}] = q[\sigma', \widehat{\mathcal{M}}]$. □

## C SMB-reduced CRN (Theorem 2)

We hereby provide the technical results to relate the notion SMB-reduced CRN, and lumped CTMC. We start providing a remark and a proposition used in the proof of Theorem 2.

*Remark 3.* Let $(S, R)$ be a CRN, $\mathcal{R}$ be an SMB, and $\mathcal{R}^\uparrow$ be its multi-set lifting. Let $\mathcal{H}^\uparrow$ be the partition induced by $\mathcal{R}^\uparrow$ on $\mathcal{MS}(\mathcal{M})$. Then, for any $(\rho_1, \rho_2) \in \mathcal{MS}(S)$ we have $\rho_1^\mathcal{R} = \rho_2^\mathcal{R}$ iff $(\rho_1, \rho_2) \in \mathcal{R}^\uparrow$. Thus, for any $\mathcal{M} \in \mathcal{H}^\uparrow$, we have $\rho_1^\mathcal{R} = \rho_2^\mathcal{R}$ for any $\rho_1, \rho_2 \in \mathcal{M}$.

In the following, we will use $\mathcal{M}^\mathcal{R}$ to denote (the unique) $\rho^\mathcal{R}$ for any $\rho \in \mathcal{M}$. Also, we will make explicit the CRN considered by each **rr** with, e.g., $\mathbf{rr}_{(S,R)}[X + Y, \mathcal{M}]$.

**Proposition 3.** *Let $(S, R)$ be a CRN, $\mathcal{R}$ be an SMB and $\mathcal{H} = S/\mathcal{R}$. Let $\mathcal{R}^\uparrow$ be the multi-set lifting of $\mathcal{R}$, and $\mathcal{H}^\uparrow$ be the partition induced by $\mathcal{R}^\uparrow$ on $\mathcal{MS}(S)$. Then, for any $X, Y \in S$ and any $\mathcal{M} \in \mathcal{H}^\uparrow$ we have*

$$(i) \ \mathbf{rr}_{(S,R)}[X + \emptyset, \mathcal{M}] = \mathbf{rr}_{(S,R)^\mathcal{R}}(X^\mathcal{R} + \emptyset, \mathcal{M}^\mathcal{R})$$

*and*

$$(ii) \ \mathbf{rr}_{(S,R)}[X, Y, \mathcal{M}] = \mathbf{rr}_{(S,R)^\mathcal{R}}(X^\mathcal{R} + Y^\mathcal{R}, \mathcal{M}^\mathcal{R})$$

*Proof.* We start addressing point $(i)$. By Definition 5 and Definition 7 we have

$$\mathbf{rr}_{(S,R)^\mathcal{R}}(X^\mathcal{R} + \emptyset, \mathcal{M}^\mathcal{R}) = \sum_{X^\mathcal{R} \xrightarrow{r} \mathcal{M}^\mathcal{R} \in R^\mathcal{R}} r = \sum_{\pi \in \mathcal{M}} \sum_{X^\mathcal{R} \xrightarrow{r} \pi \in R} r = \mathbf{rr}_{(S,R)}[X^\mathcal{R} + \emptyset, \mathcal{M}]$$

which, given that $\mathcal{R}$ is an SMB, can be further rewritten as $\mathbf{rr}_{(S,R)}[X + \emptyset, \mathcal{M}]$.

We now address point $(ii)$. We might have two cases, either $X^\mathcal{R} = Y^\mathcal{R}$, or $X^\mathcal{R} \neq Y^\mathcal{R}$. For the former case, by Definition 5 and Definition 7 we have

$$\mathbf{rr}_{(S,R)^\mathcal{R}}(X^\mathcal{R} + X^\mathcal{R}, \mathcal{M}^\mathcal{R}) = \sum_{X^\mathcal{R}+X^\mathcal{R} \xrightarrow{r} \mathcal{M}^\mathcal{R} \in R^\mathcal{R}} r = \sum_{\pi \in \mathcal{M}} \sum_{X^\mathcal{R}+X^\mathcal{R} \xrightarrow{r} \pi \in R} r = \mathbf{rr}_{(S,R)}[X^\mathcal{R} + X^\mathcal{R}, \mathcal{M}]$$

which, given that $\mathcal{R}$ is an SMB, can be further rewritten in

$$\mathbf{rr}_{(S,R)}[X + X^\mathcal{R}, \mathcal{M}] = \mathbf{rr}_{(S,R)}[X + Y, \mathcal{M}]$$

We now consider the last case: $X^\mathcal{R} \neq Y^\mathcal{R}$. By Definition 5 and Definition 7 we have

$$\mathbf{rr}_{(S,R)^\mathcal{R}}(X^\mathcal{R} + Y^\mathcal{R}, \mathcal{M}^\mathcal{R}) = \sum_{X^\mathcal{R}+Y^\mathcal{R} \xrightarrow{r} \mathcal{M}^\mathcal{R} \in R^\mathcal{R}} r = \sum_{\pi \in R} \sum_{X^\mathcal{R}+Y^\mathcal{R} \xrightarrow{r} \pi \in R} r = \mathbf{rr}_{(S,R)}[X^\mathcal{R} + Y^\mathcal{R}, \mathcal{M}]$$

which, given that $\mathcal{R}$ is an SMB, can be further rewritten in

$$\mathbf{rr}[(S,R), X + Y^\mathcal{R}, \mathcal{M}] = \mathbf{rr}[(S,R), X + Y, \mathcal{M}]$$

The proof is thus complete. $\square$

**Theorem 2 (Reduced CRNs induce lumped CTMCs).** *Let $(S, R)$ be a CRN, $\mathcal{R}$ denote an SMB and $\mathcal{H} = S/\mathcal{R}$. Further, let $\mathcal{H}^\uparrow$ denote the partition induced by $\mathcal{R}^\uparrow$ on $\mathcal{MS}(S)$. Then, for any initial population $\sigma_0$ of $(S, R)$, the underlying CTMC is such that for all $\sigma \in \mathcal{MS}(S)$ it holds that $q_{(S,R)}[\sigma, \mathcal{M}] = q_{(S^R, R^R)}(\sigma^\mathcal{R}, \mathcal{M}^\mathcal{R})$ for any $\mathcal{M} \in \mathcal{H}^\uparrow$.*

*Proof.* We may have either $\sigma \in \tilde{\mathcal{M}}$, or $\sigma \notin \tilde{\mathcal{M}}$. We start with the $\sigma \notin \tilde{\mathcal{M}}$ case, for which we also have $\sigma^\mathcal{R} \neq \tilde{\mathcal{M}}^\mathcal{R}$.

According to the proof of Theorem 1 we can rewrite $q_{(S,R)}[\sigma, \tilde{\mathcal{M}}]$ as

$$\sum_{H \in \mathcal{H}} \mathbf{rr}_{(S,R)}[X^H + \emptyset, \hat{\mathcal{M}}_H] \cdot \sigma(H) + \tag{18}$$

$$\frac{1}{2} \sum_{H \in \mathcal{H}} \sum_{\tilde{H} \in \mathcal{H}} \mathbf{rr}_{(S,R)}[X^H + Y^{\tilde{H}}, \hat{\mathcal{M}}_{H\tilde{H}}] \sum_{\substack{X \in H \\ \sigma(X) > 0}} \sigma(X) \sum_{\substack{Y \in \tilde{H} \\ (\sigma-X)(Y) > 0}} (\sigma - X)(Y) \tag{19}$$

where, as discussed in the proof of Theorem 1, $\hat{\mathcal{M}}_H$ is the block in $\mathcal{H}^\uparrow$ such that, for all $X \in H$ with $\sigma(X) > 0$, merging $\sigma - X$ with each $\pi \in \hat{\mathcal{M}}_H$ we obtain all the multi-sets $\theta \in \tilde{\mathcal{M}}$ possibly *reachable* from $\sigma$ (i.e., such that $q_{(S,R)}(\sigma, \theta)$ can be positive). Similarly for $\hat{\mathcal{M}}_{H\tilde{H}}$.

Considering instead $q_{(S,R)^\mathcal{R}}(\sigma^\mathcal{R}, \mathcal{M}^\mathcal{R})$, by Definitions 1, 2 we have

$$q_{(S,R)^\mathcal{R}}(\sigma^\mathcal{R}, \mathcal{M}^\mathcal{R}) = \sum_{\sigma^\mathcal{R} \xrightarrow{r} \mathcal{M}^\mathcal{R} \in MTS_{(S,R)^\mathcal{R}}(\sigma_0^\mathcal{R})} r = \sum_{\sigma^\mathcal{R} \xrightarrow{r} \mathcal{M}^\mathcal{R} \in out(\sigma^\mathcal{R})} r \;,$$

which in turn is equal to

$$\sum_{\substack{X \xrightarrow{r} \pi \in R^{\mathcal{R}} \\ (\sigma^{\mathcal{R}} - X) + \pi = \theta^{\mathcal{R}}}} \sigma^{\mathcal{R}}(X) \cdot r + \tag{20}$$

$$\sum_{\substack{X + X \xrightarrow{r} \pi \in R^{\mathcal{R}} \\ (\sigma^{\mathcal{R}} - (X+X)) + \pi = \theta^{\mathcal{R}}}} \sigma^{\mathcal{R}}(X) \cdot (\sigma^{\mathcal{R}} - X)(X) \cdot \frac{r}{2} + \tag{21}$$

$$\sum_{\substack{X + Y \xrightarrow{r} \pi \in R^{\mathcal{R}}, X \neq Y \\ (\sigma^{\mathcal{R}} - (X+Y)) + \pi = \theta^{\mathcal{R}}}} \sigma^{\mathcal{R}}(X) \cdot (\sigma^{\mathcal{R}} - X)(Y) \cdot r \tag{22}$$

We close the proof by showing that Equation (20) = Equation (18), and Equation (21)+Equation (22) = Equation (19). We first focus on the case Equation (20)=Equation (18). Equation (20) can be rewritten as

$$\sum_{H \in \mathcal{H}, \sigma^{\mathcal{R}}(X^H) > 0} \sigma^{\mathcal{R}}(X^H) \sum_{\substack{X^H \xrightarrow{r} \pi \in R^{\mathcal{R}} \\ (\sigma^{\mathcal{R}} - X^H) + \pi = \theta^{\mathcal{R}}}} r =$$

$$\sum_{H \in \mathcal{H}, \sigma^{\mathcal{R}}(X^H) > 0} \sigma^{\mathcal{R}}(X^H) \cdot \mathbf{rr}_{(S,R)^{\mathcal{R}}}(X^H + \emptyset, \pi_H) = \textit{(by Proposition 3)} \tag{23}$$

$$\sum_{H \in \mathcal{H}, \sigma^{\mathcal{R}}(X^H) > 0} \sigma^{\mathcal{R}}(X^H) \cdot \mathbf{rr}_{(S,R)}[X^H + \emptyset, \hat{\mathcal{M}}_H] = \textit{(by Definition 7)}$$

$$\sum_{H \in \mathcal{H}, \sigma(H) > 0} \sigma(H) \cdot \mathbf{rr}_{(S,R)}[X^H + \emptyset, \hat{\mathcal{M}}_H] \tag{24}$$

Where, to obtain Equation (23) we exploited the fact that for any $H \in \mathcal{H}$ such that $\sigma(H) > 0$ there exists one $\pi_H \in \mathcal{MS}(S)$ such that $\sigma^{\mathcal{R}} - X^H + \pi_H = \theta^{\mathcal{R}}$ [4]. Finally, $\hat{\mathcal{M}}_H$ is the block in $\mathcal{H}^\uparrow$ such that $\hat{\mathcal{M}}_H^{\mathcal{R}} = \pi_H$.

This allows us to close the case Equation (20)=Equation (18), as the considered $\hat{\mathcal{M}}_H$ in Equation (20) and Equation (24) are the same.

---

[4] If no such $\pi_H$ exists, then the **rr** has value 0.

We now address the case Equation (21)+Equation (22)=Equation (19). We first focus on Equation (21), which can be rewritten as

$$\sum_{H \in \mathcal{H}, \sigma^{\mathcal{R}}(X^H) > 1} \sigma^{\mathcal{R}}(X^H) \cdot (\sigma^{\mathcal{R}} - X^H)(X^H) \sum_{\substack{X^H + X^H \xrightarrow{r} \pi \in R^{\mathcal{R}} \\ (\sigma^{\mathcal{R}} - (X^H + X^H)) + \pi = \theta^{\mathcal{R}}}} \frac{r}{2} =$$

$$\frac{1}{2} \sum_{H \in \mathcal{H}, \sigma^{\mathcal{R}}(X^H) > 1} \sigma^{\mathcal{R}}(X^H) \cdot (\sigma^{\mathcal{R}} - X^H)(X^H) \cdot \mathbf{rr}_{(S,R)^{\mathcal{R}}}(X^H + X^H, \pi_{HH}) =$$
(25)

$$\frac{1}{2} \sum_{H \in \mathcal{H}, \sigma^{\mathcal{R}}(X^H) > 1} \sigma^{\mathcal{R}}(X^H) \cdot (\sigma^{\mathcal{R}} - X^H)(X^H) \cdot \mathbf{rr}_{(S,R)}[X^H + X^H, \hat{\mathcal{M}}_{HH}] =$$
(26)

where to obtain Equation (25) we exploited the fact that for any $H \in \mathcal{H}$ such that $\sigma^{\mathcal{R}}(X^H) > 1$ there exists one $\pi_{HH} \in \mathcal{MS}(S)$ such that $\sigma^{\mathcal{R}} - (X^H + X^H) + \pi_{HH} = \theta^{\mathcal{R}}$.[5] Instead, we obtained Equation (26) using Proposition 3. Finally, $\hat{\mathcal{M}}_{HH}$ is the block in $\mathcal{H}^\uparrow$ such that $\hat{\mathcal{M}}_{HH}^{\mathcal{R}} = \pi_{HH}$.

We now consider Equation (22), which can be rewritten as, where $1/2$ compensates the fact that each $X + Y$ is considered twice (e.g., $X = X^H, Y = X^{\tilde{H}}$, and $Y = X^H, X = X^{\tilde{H}}$)

$$\sum_{\substack{X+Y \xrightarrow{r} \pi \in R^{\mathcal{R}}, X \neq Y \\ (\sigma^{\mathcal{R}} - (X+Y)) + \pi = \theta^{\mathcal{R}}}} \sigma^{\mathcal{R}}(X) \cdot \sigma^{\mathcal{R}}(Y) \cdot r =$$

$$\frac{1}{2} \sum_{H \in \mathcal{H}, \sigma^{\mathcal{R}}(X^H) > 0} \sigma^{\mathcal{R}}(X^H) \sum_{\substack{\tilde{H} \in \mathcal{H}, H \neq \tilde{H}, \\ \sigma^{\mathcal{R}}(X^{\tilde{H}}) > 0}} \sigma^{\mathcal{R}}(X^{\tilde{H}}) \sum_{\substack{X^H + X^{\tilde{H}} \xrightarrow{r} \pi \in R^{\mathcal{R}} \\ (\sigma^{\mathcal{R}} - (X^H + X^{\tilde{H}})) + \pi = \theta^{\mathcal{R}}}} r =$$

$$\frac{1}{2} \sum_{H \in \mathcal{H}, \sigma^{\mathcal{R}}(X^H) > 0} \sigma^{\mathcal{R}}(X^H) \sum_{\substack{\tilde{H} \in \mathcal{H}, H \neq \tilde{H}, \\ \sigma^{\mathcal{R}}(X^{\tilde{H}}) > 0}} \sigma^{\mathcal{R}}(X^{\tilde{H}}) \cdot \mathbf{rr}_{(S,R)^{\mathcal{R}}}(X^H + X^{\tilde{H}}, \pi_{H\tilde{H}}) =$$
(27)

$$\frac{1}{2} \sum_{H \in \mathcal{H}, \sigma^{\mathcal{R}}(X^H) > 0} \sigma^{\mathcal{R}}(X^H) \sum_{\substack{\tilde{H} \in \mathcal{H}, H \neq \tilde{H}, \\ \sigma^{\mathcal{R}}(X^{\tilde{H}}) > 0}} \sigma^{\mathcal{R}}(X^{\tilde{H}}) \cdot \mathbf{rr}_{(S,R)}[X^H + X^{\tilde{H}}, \hat{\mathcal{M}}_{H\tilde{H}}] \quad (28)$$

where to obtain Equation (27) we exploited the fact that for any $H \in \mathcal{H}$ such that $\sigma^{\mathcal{R}}(X^H) > 0$, and any $\tilde{H}\mathcal{H}$ such that $H \neq H$ and $(\sigma^{\mathcal{R}} - X^H)(X^{\tilde{H}}) > 0$ there exists one $\pi_{H\tilde{H}} \in \mathcal{MS}(S)$ such that $\sigma^{\mathcal{R}} - (X^H + X^{\tilde{H}}) + \pi_{H\tilde{H}} = \theta^{\mathcal{R}}$.[6] Instead, in order to obtain Equation (28) we used Proposition 3. Finally, $\hat{\mathcal{M}}_{H\tilde{H}}$ is the block in $\mathcal{H}^\uparrow$ such that $\hat{\mathcal{M}}_{HH}^{\mathcal{R}} = \pi_{H\tilde{H}}$.

---

[5] If no such $\pi_{HH}$ exists, then the **rr** has value 0.

[6] If no such $\pi_{H\tilde{H}}$ exists, then the **rr** has value 0.

By putting back together the case $X = Y$ (Equation (26)) and $X \neq Y$ (Equation (28)) we obtain

$$\frac{1}{2} \sum_{\substack{H \in \mathcal{H}, \sigma^{\mathcal{R}}(X^H) > 0}} \sigma^{\mathcal{R}}(X^H) \sum_{\substack{\tilde{H} \in \mathcal{H} \\ (\sigma^{\mathcal{R}} - X^H)(X^{\tilde{H}}) > 0}} (\sigma^{\mathcal{R}} - X^H)(X^{\tilde{H}}) \cdot \mathbf{rr}_{(S,R)}[X^H + X^{\tilde{H}}, \hat{\mathcal{M}}_{H\tilde{H}}] =$$

$$\frac{1}{2} \sum_{\substack{H \in \mathcal{H}, \sigma^{\mathcal{R}}(X^H) > 0}} \sum_{\substack{\tilde{H} \in \mathcal{H} \\ (\sigma^{\mathcal{R}} - X^H)(X^{\tilde{H}}) > 0}} \mathbf{rr}_{(S,R)}[X^H + X^{\tilde{H}}, \hat{\mathcal{M}}_{H\tilde{H}}] \cdot \sigma^{\mathcal{R}}(X^H) \cdot (\sigma^{\mathcal{R}} - X^H)(X^{\tilde{H}}) =$$

$$\frac{1}{2} \sum_{H \in \mathcal{H}} \sum_{\tilde{H} \in \mathcal{H}} \mathbf{rr}_{(S,R)}[X^H + X^{\tilde{H}}, \hat{\mathcal{M}}_{H\tilde{H}}] \cdot \sum_{\substack{X \in H \\ \sigma^{\mathcal{R}}(X) > 0}} \sigma^{\mathcal{R}}(X) \sum_{\substack{Y \in \tilde{H} \\ (\sigma^{\mathcal{R}} - X)(Y) > 0}} (\sigma^{\mathcal{R}} - X)(Y)$$
(29)

where we obtained Equation (29) from the fact that, for any $H, \tilde{H} \in \mathcal{H}$, $\sigma^{\mathcal{R}}(X)$ can be positive only for $X = X^H$, and $(\sigma^{\mathcal{R}} - X)(Y)$ only for $Y = X^{\tilde{H}}$.

Finally, we close the case Equation (21)+Equation (22)=Equation (19) by showing that Equation (29) is equal to Equation (19), which directly follows from Proposition 2 (and the fact that $(\sigma, \sigma^{\mathcal{R}}) \in \mathcal{R}^{\uparrow}$, for any $\sigma \in \mathcal{MS}(S)$). The proof is thus complete. □